\documentclass{ieeeaccess}
\usepackage[sort,compress]{cite}
\usepackage{amsmath,amssymb,amsfonts}
\usepackage{algorithmic}
\usepackage{graphicx}
\usepackage{textcomp}

\usepackage{booktabs}

\usepackage{bm}
\makeatletter
\AtBeginDocument{\DeclareMathVersion{bold}
\SetSymbolFont{operators}{bold}{T1}{times}{b}{n}
\SetSymbolFont{NewLetters}{bold}{T1}{times}{b}{it}
\SetMathAlphabet{\mathrm}{bold}{T1}{times}{b}{n}
\SetMathAlphabet{\mathit}{bold}{T1}{times}{b}{it}
\SetMathAlphabet{\mathbf}{bold}{T1}{times}{b}{n}
\SetMathAlphabet{\mathtt}{bold}{OT1}{pcr}{b}{n}
\SetSymbolFont{symbols}{bold}{OMS}{cmsy}{b}{n}
\renewcommand\boldmath{\@nomath\boldmath\mathversion{bold}}}
\makeatother

\def\BibTeX{{\rm B\kern-.05em{\sc i\kern-.025em b}\kern-.08em
    T\kern-.1667em\lower.7ex\hbox{E}\kern-.125emX}}

\begin{document}

%% $$$$$$$$$$$$$$$$$$$$$$$$$$$$$$$$$$$$$$$$$$$$$$$$$$$$$$$$$$$$$$$$$$$$$$$$$$$$$

%% \history{Date of publication xxxx 00, 0000, date of current version xxxx 00, 0000.}

\history{Date of submission July 11, 2025.}

%% $$$$$$$$$$$$$$$$$$$$$$$$$$$$$$$$$$$$$$$$$$$$$$$$$$$$$$$$$$$$$$$$$$$$$$$$$$$$$

%% \doi{10.1109/ACCESS.2024.0429000}

\doi{XXX}

%% $$$$$$$$$$$$$$$$$$$$$$$$$$$$$$$$$$$$$$$$$$$$$$$$$$$$$$$$$$$$$$$$$$$$$$$$$$$$$

\title{Fast and Efficient Merge of Sorted Input Lists in Hardware Using List Offset Merge Sorters}

%% $$$$$$$$$$$$$$$$$$$$$$$$$$$$$$$$$$$$$$$$$$$$$$$$$$$$$$$$$$$$$$$$$$$$$$$$$$$$$

\author{
  \uppercase{Robert B. Kent}\authorrefmark{1}, \IEEEmembership{Life Member, IEEE}, and
  \uppercase{Marios S. Pattichis}\authorrefmark{2}, \IEEEmembership{Senior Member, IEEE}
}

%% $$$$$$$$$$$$$$$$$$$$$$$$$$$$$$$$$$$$$$$$$$$$$$$$$$$$$$$$$$$$$$$$$$$$$$$$$$$$$

\address[1]{Department of Electrical and Computer Engineering, University of New Mexico
  Albuquerque, NM 87131-0001, USA (e-mail: rkent@unm.edu)}
\address[2]{Department of Electrical and Computer Engineering, University of New Mexico,
  Albuquerque, NM 87131-0001, USA (e-mail: pattichi@unm.edu)}

%% $$$$$$$$$$$$$$$$$$$$$$$$$$$$$$$$$$$$$$$$$$$$$$$$$$$$$$$$$$$$$$$$$$$$$$$$$$$$$

\markboth
{Kent and Pattichis: Fast and Efficient Merge of Sorted Input Lists Using Hardware List Offset Merge Sorters}
{Kent and Pattichis: Fast and Efficient Merge of Sorted Input Lists Using Hardware List Offset Merge Sorters}

%% $$$$$$$$$$$$$$$$$$$$$$$$$$$$$$$$$$$$$$$$$$$$$$$$$$$$$$$$$$$$$$$$$$$$$$$$$$$$$

\corresp{Corresponding author: Robert B. Kent (e-mail: rkent@unm.edu).}

%% $$$$$$$$$$$$$$$$$$$$$$$$$$$$$$$$$$$$$$$$$$$$$$$$$$$$$$$$$$$$$$$$$$$$$$$$$$$$$

\begin{abstract}

A new set of hardware merge sort devices are introduced here, which are used to merge multiple sorted input lists of numeric values
into a single sorted output list of those values, and do so in a fast and efficient manner.
In each merge sorter, the sorted input lists are arranged in an input \mbox{2-D} setup array,
but with the order of each sorted input list offset from the order of each of the other sorted input lists.
In these new devices, called List Offset Merge Sorters (LOMS), 
a minimal set of column sort stages alternating with row sort stages process an input setup array
into a final output array, with all of the input values now in a defined sorted order.
LOMS \mbox{2-way} sorters, which merge 2 sorted input lists, 
require only 2 merge stages and are significantly faster than Kenneth Batcher's previous \mbox{state-of-the-art} \mbox{2-way} merge devices,
Bitonic Merge Sorters and \mbox{Odd-Even} Merge Sorters.  LOMS \mbox{2-way} sorters utilize the \mbox{recently-defined}
\mbox{Single-Stage} \mbox{2-way} Merge Sorters (S2MS) in their first stage.  Both LOMS and S2MS devices can merge any mixture of input list sizes,
while Batcher's merge sorters are difficult to design unless the 2 input lists are equal, and a \mbox{power-of-2}.
By themselves, S2MS devices are the fastest \mbox{2-way} merge sorters when implemented in this study's target FPGA devices,
but they tend to use a large number of LUT resources.
LOMS \mbox{2-way} devices use fewer resources than comparable S2MS devices, enabling some large LOMS devices
to be implemented in a given FPGA when comparable S2MS devices cannot fit in that FPGA.
In addition, the study's smallest \mbox{2-way} LOMS device uses fewer resources than the comparable Bitonic merge sorter, even though it is faster.
A List Offset \mbox{2-way} sorter merges 2 lists, each with 32 \mbox{32-bit} values, into a sorted output list of those 64 values in 2.24 nS,
a speedup of 2.63 versus a comparable Batcher \mbox{2-way} merge device.
A LOMS \mbox{3-way} merge sorter, merging 3 sorted input lists, each with 7 values,
fully merges the 21 \mbox{32-bit} values in 3.4 nS, a speedup of 1.36 versus the comparable \mbox{state-of-the-art} \mbox{3-way} merge device.

\end{abstract}

%% $$$$$$$$$$$$$$$$$$$$$$$$$$$$$$$$$$$$$$$$$$$$$$$$$$$$$$$$$$$$$$$$$$$$$$$$$$$$$


\titlepgskip=-21pt

\maketitle

%% $$$$$$$$$$$$$$$$$$$$$$$$$$$$$$$$$$$$$$$$$$$$$$$$$$$$$$$$$$$$$$$$$$$$$$$$$$$$$

\section{Nomenclature}

\label{nomenclature}

\vspace{2pt}

\noindent
\textbf{k-way merge sort}\\
\hangindent 0.35cm
Merge of k sorted inputs lists into a sorted output list.

\noindent
\textbf{LOMS}\\
\hangindent 0.35cm
List Offset Merge Sort

\noindent
\textbf{S2MS}\\
\hangindent 0.35cm
Single-Stage 2-way Merge Sort

\noindent
\textbf{OEMS}\\
\hangindent 0.35cm
Odd-Even Merge Sort

\noindent
\textbf{BiMS}\\
\hangindent 0.35cm
Bitonic Merge Sort

\noindent
\textbf{MWMS}\\
\hangindent 0.35cm
Multiway Merge Sort

\noindent
\textbf{FPGA}\\
\hangindent 0.35cm
Field-Programmable Gate Array

\noindent
\textbf{2insLUT}\\
\hangindent 0.35cm
FPGA S2MS and LOMS sorters with 2 input bits per LUT

\noindent
\textbf{4insLUT}\\
\hangindent 0.35cm
FPGA S2MS and LOMS sorters with 4 input bits per LUT

\noindent
\textbf{UP-x/DN-y}\\
\hangindent 0.35cm
A 2-way merge of sorted UP input list with x values,\\
sorted DN input list with y values

\noindent
\textbf{Xc\_Yr}\\
\hangindent 0.35cm
An X-way merge of X sorted input lists, each list having Y values,
placed into an X column, Y row setup array.

%% $$$$$$$$$$$$$$$$$$$$$$$$$$$$$$$$$$$$$$$$$$$$$$$$$$$$$$$$$$$$$$$$$$$$$$$$$$$$$

\section{Introduction}

\label{sec:introduction}

\PARstart{N}{umber} sorting in hardware has emphasized using algorithms that take advantage of the parallel processing
and pipelining that hardware devices typically provide.  The dominant 2 methodologies for implementing hardware sort
have been Kenneth Batcher's \mbox{Odd-Even} Merge Sort and Bitonic Merge Sort \cite{Batcher_both}.
When starting from an unsorted list of N values, each methodology begins by using \mbox{2-sorters}
in parallel to produce N{/}2 sorted pairs of values.  After this first operation on an unsorted list,
multistage \mbox{2-way} merge sequences are used, once again in parallel,
to merge sets of 2 sorted inputs lists to produce larger sorted output lists.

Methods for designing \mbox{Single-Stage} \mbox{2-way} Merge (S2SM) sorters
have recently been introduced \cite{s2sm_Asilomar}\cite{s2sm_patent}.
The \mbox{Single-Stage} \mbox{2-way} Merge sorters constructed using these methods merge 2 sorted input lists
faster than Batcher's multistage \mbox{2-way} merge devices, but tend to use significantly more FPGA hardware resources.

In order to reduce hardware resource usage, while still enabling fast operation,
new List Offset Merge Sorters are introduced here for \mbox{2-way} merge sort.
The new List Offset \mbox{2-way} merge devices are \mbox{2-stage} devices.
They are slower than \mbox{single-stage} \mbox{2-way} merge devices, but use fewer resources,
and they are still faster than Batcher's \mbox{2-way} merge devices,
particularly as the size of the sorted input lists increases.  Since \mbox{2-way} LOMS devices use fewer resources than S2MS devices,
some larger LOMS devices can be constructed in a target FPGA, while the comparable S2MS devices are too large to be placed.

List Offset methods can also be used for \mbox{k-way} merge of k sorted inputs lists, where k{>}2.
When k{=}3, it takes List Offset 3 stages to fully merge the 3 sorted input lists.
If the number of values in the sorted inputs are equal and odd,
it only takes 2 List Offset stages to determine the output list median.
An existing set of \mbox{k-way} merge devices \cite{us_access_multiway_2022}\cite{multiway_merge_patent}
are considered here to be the \mbox{state-of-the-art} for hardware \mbox{k-way} merge,
and it will be shown that List Offset \mbox{3-way} merge devices are considerably faster
than the comparable \mbox{state-of-the-art} \mbox{3-way} merge devices.

All of the hardware merge sorters discussed here are oblivious, \mbox{hard-wired}.
Their design and operation are fixed, and operate on all sets of sorted input values in the same manner.
Although designed for hardware, these oblivious processes can be implemented in software for safety or security reasons.

The hardware merge sorters characterized in this study have been constructed in 2 FPGA products,
from 2 distinctly different FPGA families.
The 2 products are the AMD Kintex Ultrascale+ \mbox{xcku5p-ffva676-3-e}
and the AMD Versal Prime \mbox{xcvm1102-sfva784-2HP-i-S}.
All designs were synthesized for these 2 products using AMD's 2024.2 Vivado tool.
Neither product requires a license in order to run on Vivado 2024.2.

After covering the background of \mbox{2-way} and \mbox{k-way} merge sort, a detailed description of the design and operation
of \mbox{2-way} List Offset Merge Sorters is presented in Section~\ref{sec:list_offset_2_way_merge}.
Next, \mbox{k-way} LOMS devices are covered in

Section~\ref{sec:list_offset_k_way_merge}, particularly focusing on \mbox{3-way} merge.
Section~\ref{sec:list_offset_characterization} discusses specifics of the \mbox{2-way} and \mbox{3-way} merge sorter designs
used for this study's data gathering and analysis.
The speed and resource usage results for the various merge sorters are then presented in Section~\ref{sec:top_results}.

%% $$$$$$$$$$$$$$$$$$$$$$$$$$$$$$$$$$$$$$$$$$$$$$$$$$$$$$$$$$$$$$$$$$$$$$$$$$$$$

\section{Background}

\label{sec:background}

Hardware sorters constructed with either of Kenneth Batcher's 2 classic algorithms, 
\mbox{Odd-Even} Merge Sort and Bitonic Merge Sort \cite{Batcher_both}, feature parallel sort operations in each stage of a sorting
process, and the stages can be pipelined as well.
Although Batcher's sorters can be clocked and pipelined, the algorithms themselves are combinatorial.  The input port values can propagate
to the output ports through logic structures, like FPGA Look Up Tables (LUTs), without need for clocking.
His 2 sorting algorithms continue to be the most popular methods for designing and constructing hardware sorting devices
\cite{Wisc_N_to_4,Zuluaga_spiral,Zuluaga_streaming,chen_prasanna,no_feedback,ferger_and_blott,rths,flims,Parallel_merge_sorter_patent_2023,Oh_streaming_2024}
.
Since Batcher's merge sorting algorithms produce an oblivious sort, they can be used in software designed for safety and security
\cite{Ngai_2024}.

As mentioned previously, when working on an unsorted list of values, the first stage for each of Batcher's processes consists
of using \mbox{2-sorters} in parallel to change the unsorted list into a group of sorted lists, each with 2 values.  Any stage
after the first stage is part of a multistage \mbox{2-way} merge sequence, in which groups of 2 sorted input lists are merged into a
single sorted output list.

Two stages in a Batcher process are required for a \mbox{2-way} merge of 2 sorted input lists, each with 2 values,
into a single sorted output list of 4 values.  Each doubling of list size adds one additional stage to the sequence
required to implement the \mbox{2-way} merge.

An earlier work \cite{Adas_2007} did define a \mbox{single-stage} \mbox{2-way} merge sorter.
The device featured an \mbox{in-place} set of m registers, and n sorted input values were to be merged
with the m values.  In each clock cycle, the highest m values of the m+n total inputs were then stored into the m registers,
and the lowest n values were ignored and lost. As this was an \mbox{in-place} device, the sorting process itself is always clocked.
No details were given concerning how the m output multiplexers were constructed and controlled.

A new system has recently been defined for designing, building, and operating \mbox{Single-Stage} \mbox{2-way} Merge Sorters~
\cite{s2sm_Asilomar}\cite{s2sm_patent}.
Like Batcher's sorters, these S2SM sorters are defined as combinatorial.  Since they are \mbox{single-stage} devices, 
their internal operation will not need to be clocked or pipelined.  However, if the \mbox{single-stage} operation does not complete
within one cycle of a fast clock, their internal operation can be clocked and pipelined.
For the \mbox{2-way} List Offset Merge Sorters defined in this study, which are \mbox{2-stage} merge sorters,
it is assumed that the first stage will be implemented by 2 or more of these S2SM devices,
operating in parallel, sorting each of the columns of \mbox{2-D} arrays.

Merge sort of more than 2 lists, \mbox{k-way} merge, with k{>}2, has not been emphasized as much as \mbox{2-way} merge.
One early paper defining a \mbox{k-way} merge sort system was \cite{sloping_and_shaking}.
Papers dealing with \mbox{3-way} merge tend to focus on
finding the median of a 3x3 set of values for image noise reduction \cite{median_filter_core_2024}.  A general system for designing
and implementing \mbox{k-way} merge sorters is considered here to be the \mbox{state-of-the-art} for \mbox{k-way} merge
\cite{us_access_multiway_2022}\cite{multiway_merge_patent}, and one of its \mbox{3-way} merge sorters will be used to assess
the performance of the comparable \mbox{3-way} LOMS sorter.

Hardware \mbox{2-sorters}, which sort 2 unsorted values, are easily constructed in FPGAs, and have long been the default devices
used in \mbox{2-way} merge sort.  In \mbox{this work}, \mbox{single-stage} \mbox{N-sorters}, which sort an unsorted list of N values with N{>}2
\cite{us_1}\cite{us_2},
are used for the row sorters of \mbox{2-way} LOMS devices with more than 2 columns, multiway merge \mbox{3-way},
and LOMS \mbox{3-way} merge sorters.

%% $$$$$$$$$$$$$$$$$$$$$$$$$$$$$$$$$$$$$$$$$$$$$$$$$$$$$$$$$$$$$$$$$$$$$$$$$$$$$

\section{List Offset 2-way Merge}

\label{sec:list_offset_2_way_merge}

For any List Offset merge, the sorted input lists are mapped into a \mbox{2-D} setup array in the target hardware,
with the ordering of each list offset from that of previous lists.
After an input setup array is constructed, column sort stages alternate with row sort stages,
until the final \mbox{2-D} array is in sorted order, after which the array values are mapped into the
sorted output list.
The number of required alternating column sort/row sort stages depends on the number of sorted input lists to be merged.

In this section, List Offset \mbox{2-way} merge sort is introduced, in which 2 sorted input lists are merged into a single sorted output list.
List Offset \mbox{2-way} merge requires only 2 stages, a full column sort of each column, followed by a full row sort of each row.

\begin{figure}[h]
  \centering
  \includegraphics[width=0.9\linewidth]{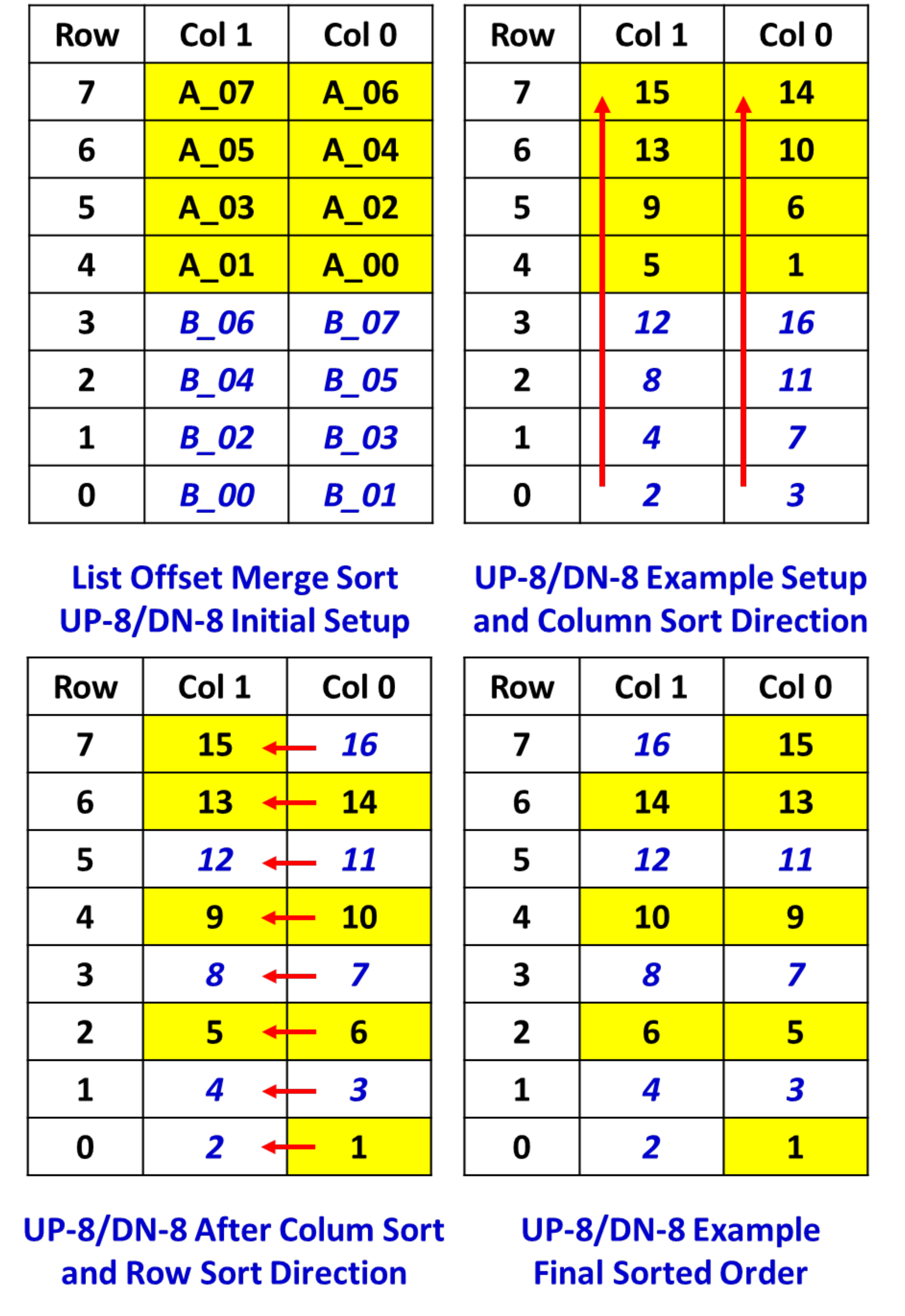}
  \caption{List Offset UP-8/DN-8 Example Setup and 2-stage Operation.}
  \label{fig:A_up_8_dn_8_4_arrays}
\end{figure}

The upper left array of Fig.~\ref{fig:A_up_8_dn_8_4_arrays} shows the \mbox{2-column} input setup array
for an UP-8/DN-8 LOMS merge, with 2 sorted input lists, each with 8 values.  The A UP list values are mapped into the top 4 rows of the array.
The A list is sorted from the max at A\_07 to the min at A\_00.  The highest A values are in the top row down to the lowest values
in the 4th row down.  Each row has its max value in Col 1, the leftmost column, and its min value in Col 0, the rightmost column.

The B DN list values are mapped into the lowest 4 rows of the array.  The B list values range from the max at B\_07 to the min at B\_00.
The highest B values are in Row 3, the highest B row, down to the lowest B values in Row 0.
In each B row, the sorted order is opposite to the A list sorted row order.
The highest row value is in Col 0, the rightmost column, and the lowest row values in Col 1, the leftmost column.

\mbox{Example A and} B sorted list values are shown in upper right array in Fig.~\ref{fig:A_up_8_dn_8_4_arrays},
as well arrows in each column, indicating that the first stage operations consist of separate, parallel sorts of each column.
The example values after the column sorts are shown in the lower left Fig.~\ref{fig:A_up_8_dn_8_4_arrays} array,
with arrows in each row indicating the separate, parallel row sorts for the second stage operations.

The lower right Fig.~\ref{fig:A_up_8_dn_8_4_arrays} array shows the final sorted order after the \mbox{2-stage} sequence.
The max of each row is in Col 1, the leftmost column, and the min of each row is in Col 0, the rightmost column.

Any \mbox{2-way} merge 2 column array constructed in this manner only requires the column sort and row sort stages,
after which the array is in final sorted order.  The A/B list lengths do not need to be equal, and each length can be either odd or even.

\begin{figure}[h]
  \centering
  \includegraphics[width=0.9\linewidth]{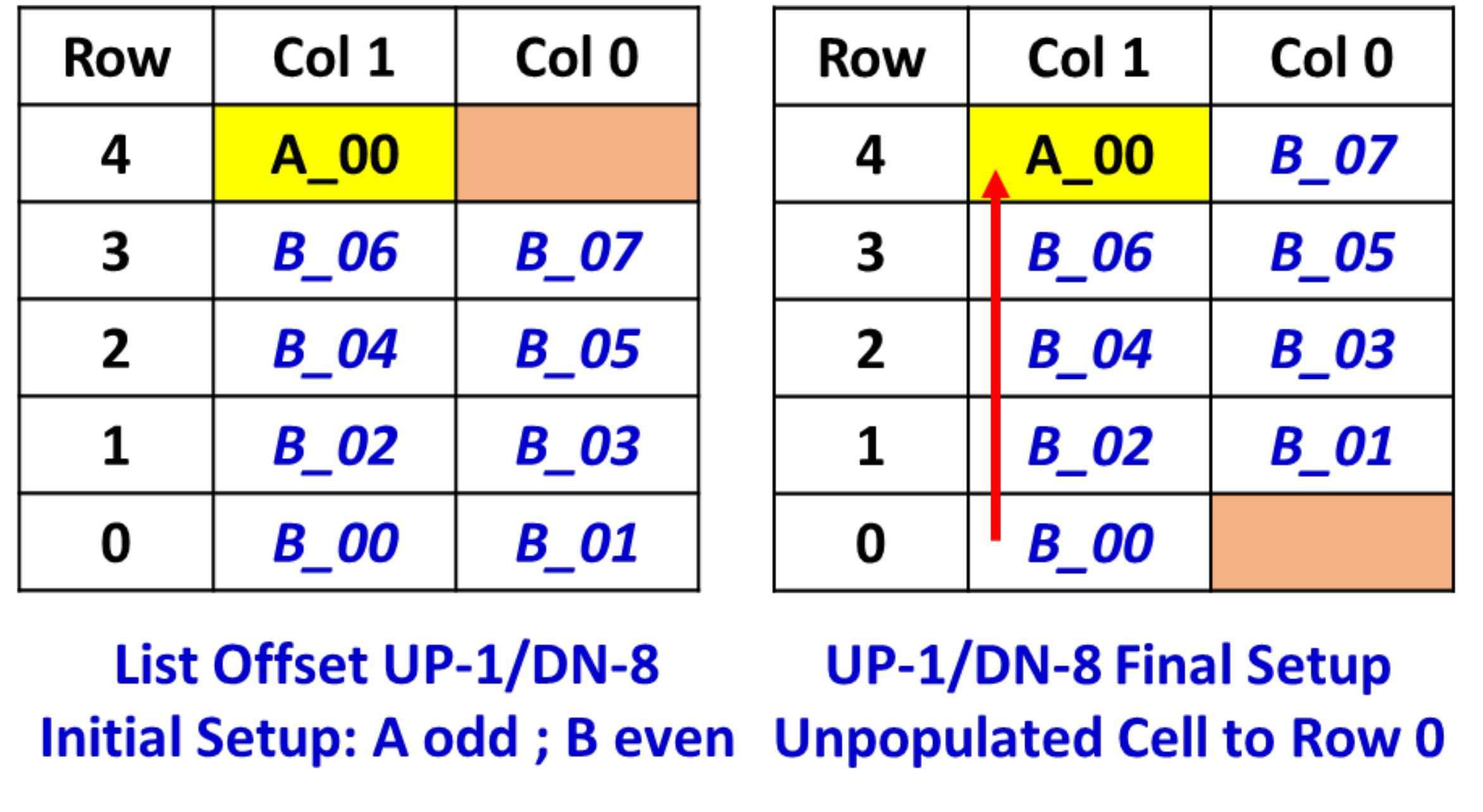}
  \caption{List Offset Initial and Final UP-1/DN-8 Setups; Sort only Col 1.}
  \label{fig:A_odd_1_B_even_8}
\end{figure}

Fig.~\ref{fig:A_odd_1_B_even_8} shows the setup for an \mbox{UP-1/DN-8} sorter, with the A UP list odd and the B DN list even.
The left Fig.~\ref{fig:A_odd_1_B_even_8} array shows the initial setup, in which there is an unpopulated cell in \mbox{Col 0},
above the DN value rows.  A List Offset unpopulated cell needs to move down to Row 0 in the final setup, and this is what happens
in the Fig.~\ref{fig:A_odd_1_B_even_8} right array.  In the Fig.~\ref{fig:A_odd_1_B_even_8} right array, there are no
A list values in Col 0, the B list values in Col 0 are already in column sorted order,
so Col 0 does not need a column sort in the List Offset first stage.

In the first column sort, the overall min value, the min of A\_00 and B\_00 will end up in Row 0,
with no subsequent row sort needed for this bottom row.
All other rows in the array will require a row sort to complete the \mbox{2-way} merge sort.

\begin{figure}[h]
  \centering
  \includegraphics[width=0.9\linewidth]{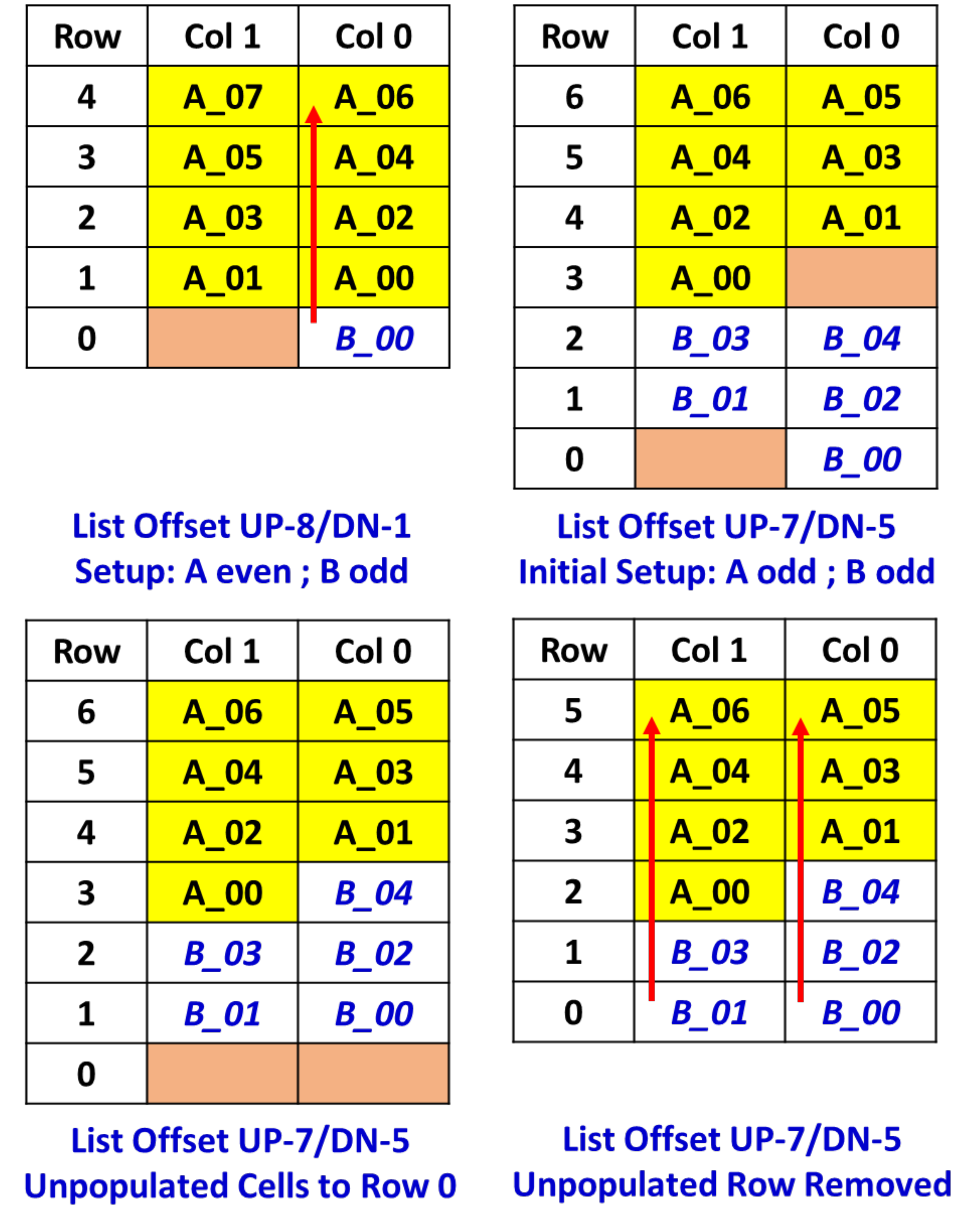}
  \caption{List Offset UP-8/DN-1, UP-7/DN-5 Setups, Column Sort arrows.}
  \label{fig:A_even_8_B_odd_1}
\end{figure}

No adjustment is needed for \mbox{UP-8/DN-1} sorter initial setup shown in the upper left array of 
Fig.~\ref{fig:A_even_8_B_odd_1}, with the A list even and the B list odd.
The only unpopulated cell in the initial setup is from the B list, and it is already in Row 0.
Only \mbox{Col 0} requires a first stage column sort, as the values in \mbox{Col 1} are all from the same list,
and are therefore already in sorted order.

Once again, after the first stage column sort, the overall min value, the min of A\_00 and B\_00 will already be in
the only populated cell in Row 0. All other rows will require a second stage row sort.

The upper right array in Fig.~\ref{fig:A_even_8_B_odd_1} shows the initial setup for an \mbox{UP-7/DN-5} List Offset sorter,
in which both lists have an odd number of values.  The unpopulated cell above the B list values needs to slide
down to Row 0, and this is what happens, as shown in the lower left array of
Fig.~\ref{fig:A_even_8_B_odd_1}.
This leaves Row 0 fully unpopulated, so it is removed for the final setup, shown in the 
Fig.~\ref{fig:A_even_8_B_odd_1}
lower right array.

By taking a closer look at the setup arrays presented in this section, it starts to become clear how the Stage 1 column sorts
are able to move the various input values into their correct row.  In each setup array, the 2 highest values, either the
top 2 values in a single list, or the max values of the separate lists, will always move to the top row during the Stage 1
column sorts.  Likewise, the min values will always move to their final bottom row during Stage 1.  This phenomenon can be studied
for the values moving to other rows, and it will then become clear why the List Offset setup arrays always enable values
to go their final row locations during the Stage 1 column sort.

It should be clear from the \mbox{2-way} merge figures in this section that any 2 sorted lists can be set up
and merged in 2 stages using List Offset merge sorters.  It should also be noted that any column sort
shown in the figures can be performed by an S2MS \mbox{Single-Stage} \mbox{2-way} Merge Sort,
as both the A and B values are in sorted order in each column.

Although all of the figure arrays so far use 2 columns, a List Offset \mbox{2-way} merge array can have more than 2 columns.
For a given output list size, increasing the number of array columns
reduces the size of the column sorters operating in parallel in the 1st stage,
but increases the size of the \mbox{single-stage} parallel row sorters in the 2nd stage.

When there are more than 2 columns in a \mbox{2-way} List Offset \mbox{2-D} array,
the 2 input lists are mapped to the initial array in a manner similar to that when using 2 columns.
The largest values are placed in the highest rows occupied by a list with the lowest values placed in the lowest rows.
The values in an A UP list row range from the max on the left to the min on the right.
The B DN list row order is reversed, ranging from the max on the right edge of the row to the min on the left row edge.

\begin{figure}[h]
  \centering
  \includegraphics[width=\linewidth]{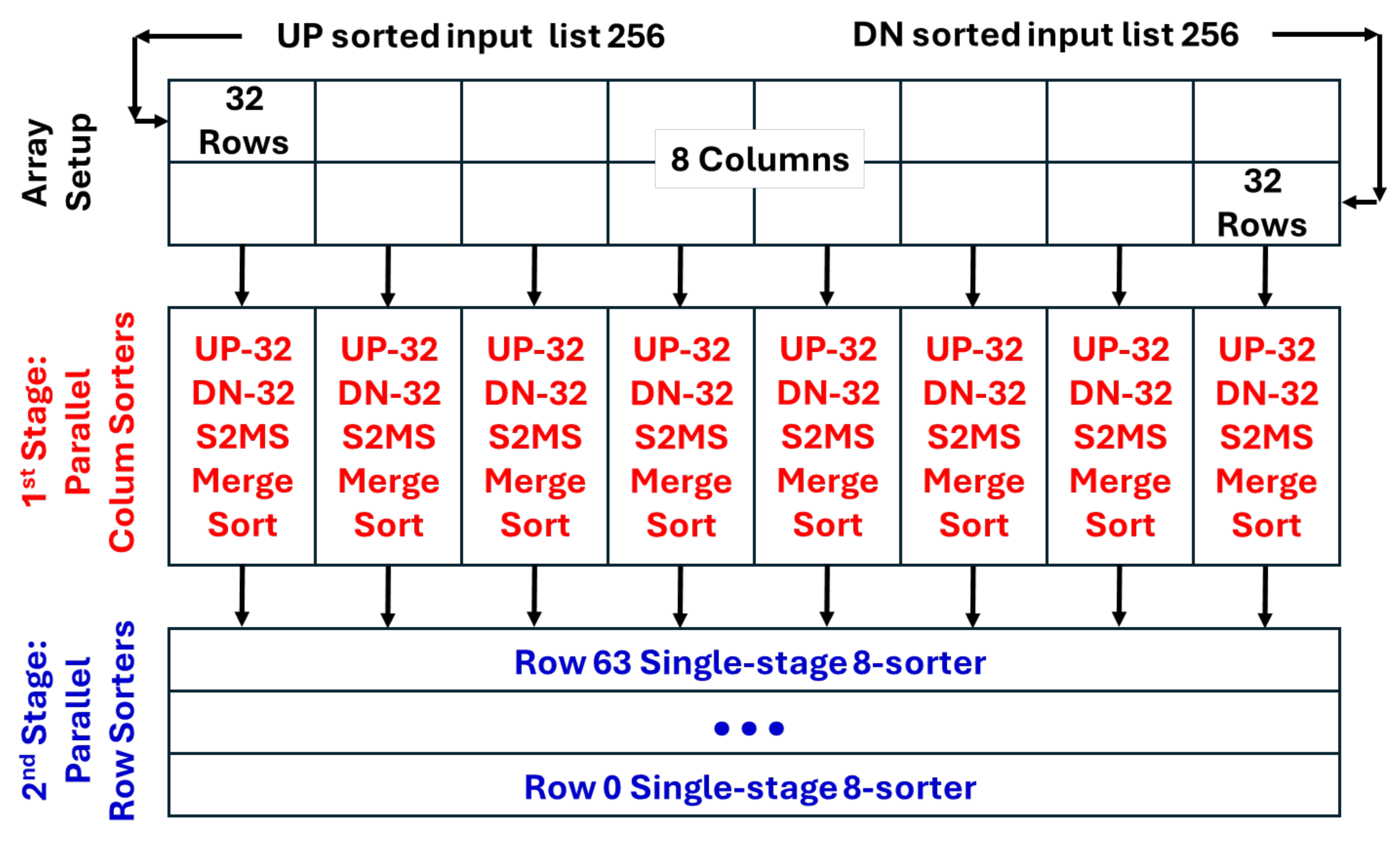}
  \caption{8-column UP-256/DN-256 2-way List Offset Merge Sorter.}
  \label{fig:orms_8_columns}
\end{figure}

A \mbox{top-level} diagram for an \mbox{8-column} \mbox{UP-256/DN-256} List Offset Merge Sorter is shown in
Fig.~\ref{fig:orms_8_columns}.  It uses 8 S2MS \mbox{32-UP/32-DN} merge sorters as the column sorters.
This is the largest List Offset Merge Sorter constructed and characterized in this study.
Because of the large resource usage of S2MS devices,
an S2MS \mbox{UP-256/DN-256} merge sorter could not be fit into either of the 2 target FPGA products.

%% $$$$$$$$$$$$$$$$$$$$$$$$$$$$$$$$$$$$$$$$$$$$$$$$$$$$$$$$$$$$$$$$$$$$$$$$$$$$$

\section{List Offset k-way Merge}

\label{sec:list_offset_k_way_merge}

List Offset \mbox{2-D} arrays and the processing on those arrays can be used for fast and efficient \mbox{k-way} merge,
merging k{>}2 sorted inputs lists.  For example, a \mbox{3-way} merge is fully sorted after 3 stages,
but median and other values can be determined after only 2 stages, the full column sort stage followed by the full row sort stage.
An example of LOMS 3-way merge operation is shown in Section~\ref{sec:list_offset_3_way_merge} below.
Section~\ref{sec:list_offset_more_than_3_way_merge} then discusses LOMS devices which merge more than 3 sorted input lists.

%% $$$$$$$$$$$$$$$$$$$$$$$$$$$$$$$$$$$$$$$$$$$$$$$$$$$$$$$$$$$$$$$$$$$$$$$$$$$$$

\subsection{List Offset 3-way Merge}

\label{sec:list_offset_3_way_merge}

\begin{figure}[h]
  \centering
  \includegraphics[width=1.0\linewidth]{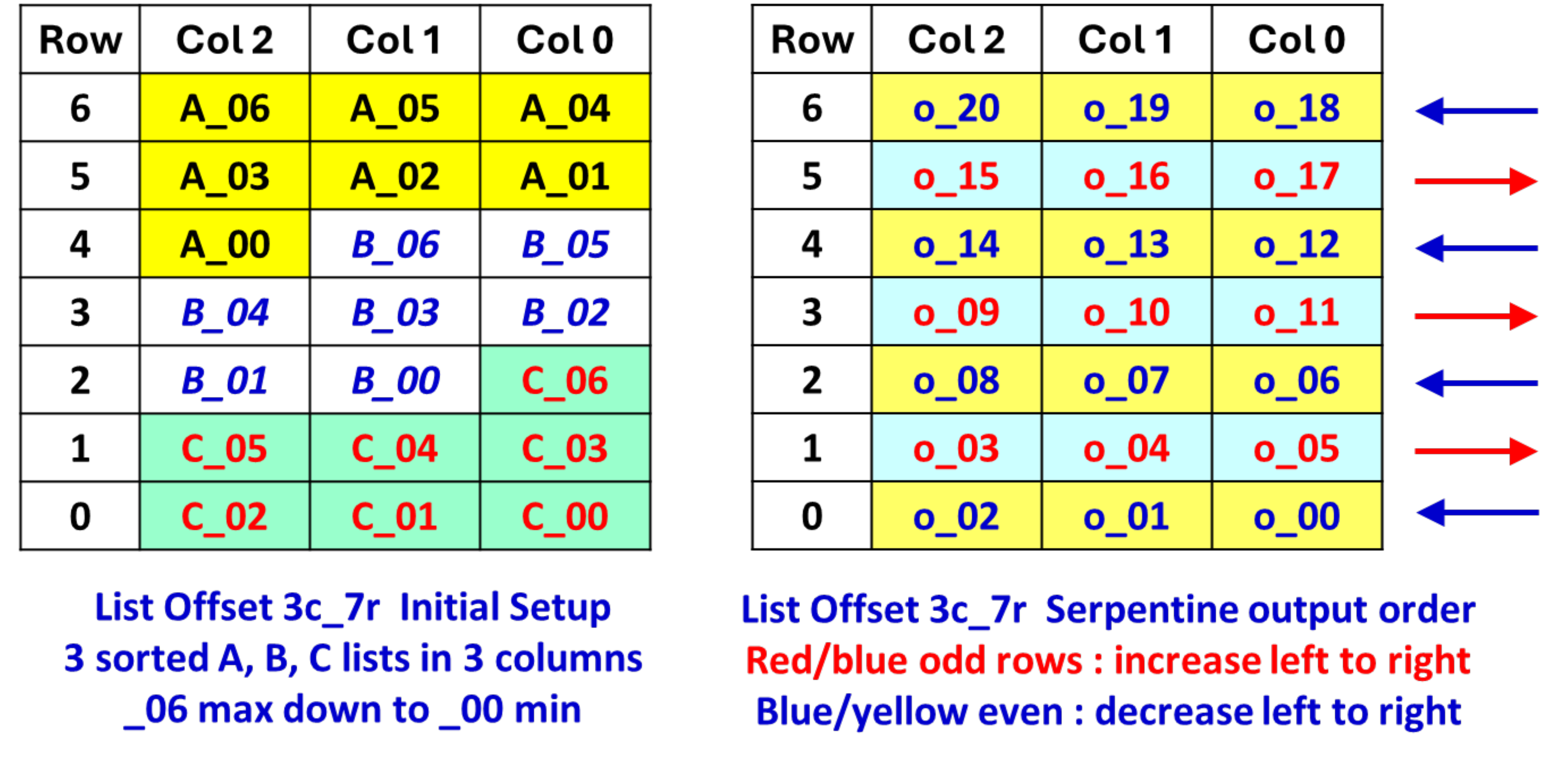}
  \caption{3c\_7r 3-way 3-column setup; Sorted output order.}
  \label{fig:3c_7r_setup_and_outputs}
\end{figure}

The 3-column left array in Fig.~\ref{fig:3c_7r_setup_and_outputs} shows the array setup, prior to processing,
for a 3c\_7r 3-way merge.  It should be clear that the values from each of the 3 A, B, and C sorted input lists
are offset from the those of the other 2 lists.  Appendix~\ref{app:3_way_merge_setup_design} shows how this LOMS setup array
is constructed, and is a template for constructing any k-column LOMS setup array containing k sorted input lists.

The right Fig.~\ref{fig:3c_7r_setup_and_outputs} array shows the final output sorted order that is defined for this 3c\_7r LOMS.
This is a serpentine sorted order, with the sorted order of each row reversed from the rows above and below that row.  This
serpentine order is defined for all \mbox{k-column}, \mbox{k-way} merge devices with k${\ge}$3,
as it enables the simple alternating column sort/row sort merge stages after Stage 2.

\begin{figure}[h]
  \centering
  \includegraphics[width=1.0\linewidth]{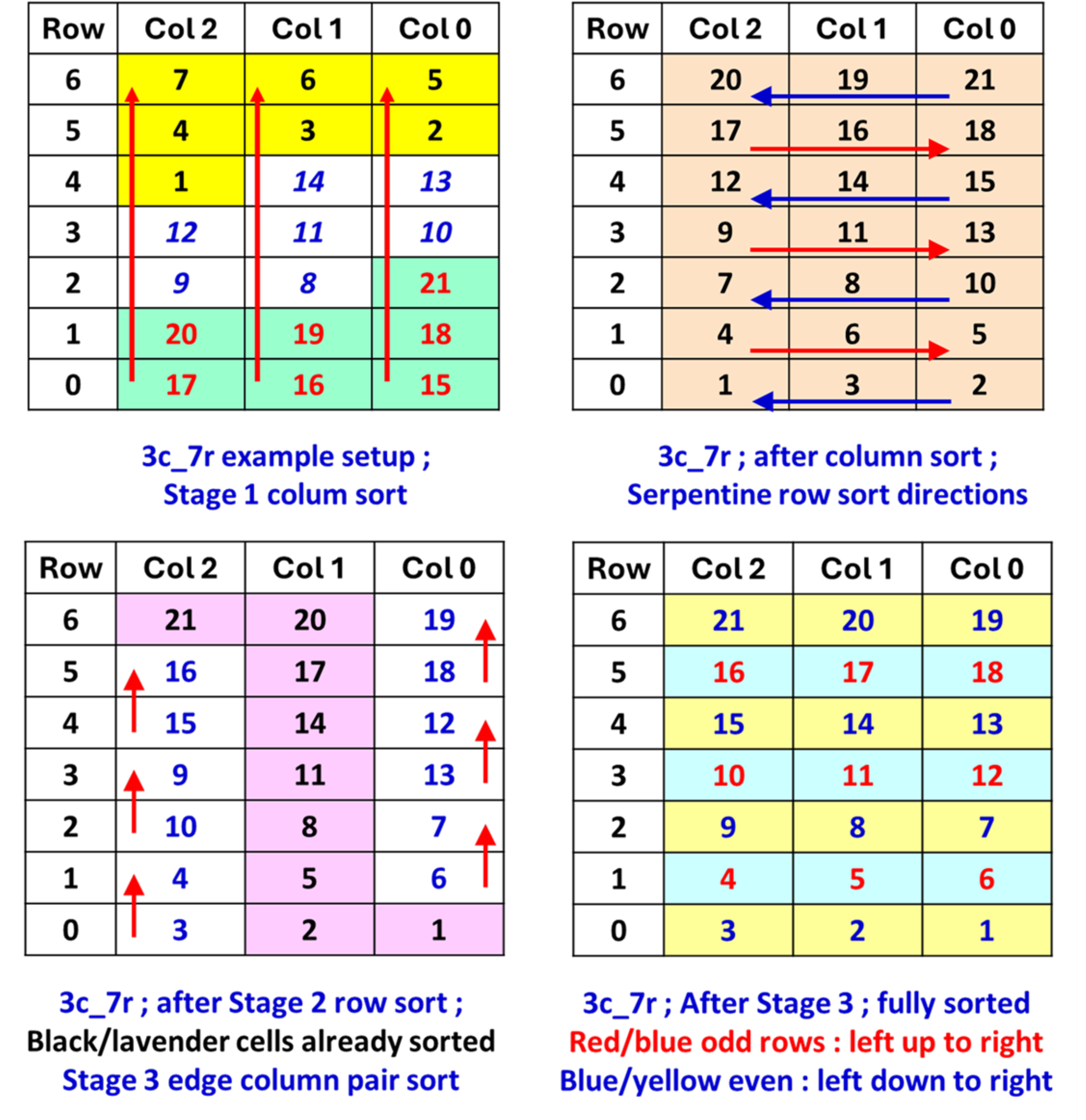}
  \caption{3c\_7r 3-stage 3-way merge.}
  \label{fig:3c_7r_3_stage_3_way_merge}
\end{figure}

The upper left Fig.~\ref{fig:3c_7r_3_stage_3_way_merge} array shows a 3c\_7r setup array, like the
left Fig.~\ref{fig:3c_7r_setup_and_outputs} array, but  with numeric A, B, and C list values.
The vertical red arrows in this array indicate the Stage 1 full column sorts.  This particular example appears to be a worst
case for LOMS merge sort.  The highest of the 21 values are in the bottom rows, but need to move to the top of the array,
and in the correct final order columns.  Likewise, the lowest of the 21 values are at the top of the setup array, but
need to move to array bottom, and to the correct column locations.

The  Fig.~\ref{fig:3c_7r_3_stage_3_way_merge} upper right array shows the array values after the Stage 1 column sort.
This upper right array also shows the alternating directions of the parallel Stage 2 row sorts.

The Fig.~\ref{fig:3c_7r_3_stage_3_way_merge} lower left array shows the array values after the Stage 2 serpentine row sort.
The values in the cells with lavender fill are already in their final order locations, so those cells need no Stage 3
column sort.  The median of the 21 input values is found at Row 3 Col 1, and it has already been determined after Stage 2.
The column sort operations in Stage 3 sort pairs of values in the edge columns Col 2 and Col 0.

The final sorted order of the 21 values is shown in the lower right Fig.~\ref{fig:3c_7r_3_stage_3_way_merge} array.
The even rows, with values nonincreasing from Col 2 to Col 0, are shown in blue text, with yellow cell fill.
The odd rows, with values nondecreasing from Col 2 to Col 0, are shown in red text, with light blue cell fill.

%% $$$$$$$$$$$$$$$$$$$$$$$$$$$$$$$$$$$$$$$$$$$$$$$$$$$$$$$$$$$$$$$$$$$$$$$$$$$$$

\subsection{List Offset Merge of More Than 3 Input Lists}

\label{sec:list_offset_more_than_3_way_merge}

The design of setup arrays for LOMS \mbox{k-way} merge devices, with k{>}3,
is similar to the design of LOMS \mbox{3-way} setup arrays.
A \mbox{k-way} merge setup array is constructed in a \mbox{k-column} array,
and follows the steps shown in Appendix~\ref{app:3_way_merge_setup_design} for a \mbox{3-way} merge.

As always, the first 2 \mbox{k-way} LOMS stages consist of parallel full column sorts followed by parallel full for row sorts.
After the first 2 stages, column sort stages alternate with row sort stages until the values in the array are always in the correct sorted order.
However, after the first 2 stages, the column sorts and row sorts may not sort full columns or full rows,
and some columns and rows may not be sorted at all.
For example, for the 3rd column sort stage of the \mbox{3-way} merge shown in the lower left Fig.~\ref{fig:3c_7r_3_stage_3_way_merge} array,
Col 1 values are not sorted at all, and the Col 2 and Col 0 sort operations consist of sorting pairs of values in those columns.

Table~\ref{tab:phases_vs_nlists} lists the number of alternating column and row sorts required to fully sort a List Offset
design with k sorted input lists.  Specific details on the design and operations of List Offset Merge Sorters
which merge more than 3 lists are not presented here.

\begin{table}[h]
  \caption{Total Column/Row sorts required for a k-way merge.}
  \label{tab:phases_vs_nlists}
    \centering
      \begin{tabular}{cccccccc}
        \toprule
%% <<<<<<<<<<<<<<<<<<<<<<<<<<<<<<<<<<<<<<<<<<<<<<<<<<<<<<<<<<<<<<<<<<<<<<<<<<<<<<<<<<<<<<<<<<<<<<<<<<<<<<<<<<<<<<<<<<<<<<<<<<<<<<
~ k No.     &           &       &        &       &        &        & Total  \\
~ Sorted    &  1st      & 1st   &  2nd   &  2nd  &  3rd   &  3rd   & Col \&  \\
~ Input     &  Col      & Row   &  Col   &  Row  &  Col   &  Row   & Row    \\
~ Lists     &  Sort     & Sort  &  Sort  & Sort  &  Sort  & Sort   & Sorts  \\ \midrule
 %% <<<<<<<<<<<<<<<<<<<<<<<<<<<<<<<<<<<<<<<<<<<<<<<<<<<<<<<<<<<<<<<<<<<<<<<<<<<<<<<<<<<<<<<<<<<<<<<<<<<<<<<<<<<<<<<<<<<<<<<<<<<<<
    2       &   \checkmark       &   \checkmark   &        &       &        &        &   2    \\  
    3       &   \checkmark       &   \checkmark   &   \checkmark    &       &        &        &   3    \\  
    4       &   \checkmark       &   \checkmark   &   \checkmark    &   \checkmark   &        &        &   4    \\  
    5       &   \checkmark       &   \checkmark   &   \checkmark    &   \checkmark   &        &        &   4    \\  
    6       &   \checkmark       &   \checkmark   &   \checkmark    &   \checkmark   &   \checkmark    &        &   5    \\  
    7       &   \checkmark       &   \checkmark   &   \checkmark    &   \checkmark   &   \checkmark    &   \checkmark    &   6    \\  
    {...}   &   \checkmark       &   \checkmark   &   \checkmark    &   \checkmark   &   \checkmark    &   \checkmark    &   6    \\  
    14      &   \checkmark       &   \checkmark   &   \checkmark    &   \checkmark   &   \checkmark    &   \checkmark    &   6    \\  \bottomrule
%% <<<<<<<<<<<<<<<<<<<<<<<<<<<<<<<<<<<<<<<<<<<<<<<<<<<<<<<<<<<<<<<<<<<<<<<<<<<<<<<<<<<<<<<<<<<<<<<<<<<<<<<<<<<<<<<<<<<<<<<<<<<<<<
      \end{tabular}
\end{table}

%% $$$$$$$$$$$$$$$$$$$$$$$$$$$$$$$$$$$$$$$$$$$$$$$$$$$$$$$$$$$$$$$$$$$$$$$$$$$$$

\section{LOMS FPGA Design Methodology}

\label{sec:list_offset_characterization}

In order to analyze the performance of various List Offset Merge Sorters, they have been 
constructed in 2 FPGA devices, from 2 separate FPGA device families, and then compared to
\mbox{same-size} \mbox{state-of-the-art} merge sorters, also constructed in the same 2 FPGAs.

As mentioned previously, the two FPGA products are the Kintex Ultrascale+ \mbox{xcku5p-ffva676-3-e} 
and the Versal Prime \mbox{xcvm1102-sfva784-2HP-i-S}.
Merge sorters were constructed for both \mbox{8-bit} and \mbox{32-bit} unsigned integer values.

For \mbox{2-way} merge sort, the devices previously considered to be \mbox{state-of-the-art} are those constructed with Kenneth Batcher's
\mbox{Odd-Even} Merge Sort or Bitonic Merge Sort. The input and output lists sizes used for the \mbox{2-way} merge sort
analysis here are integer \mbox{powers-of-2}, as Batcher's devices are most efficient for these list sizes,
and difficult to design for other list sizes.
LOMS and S2MS \mbox{2-way} merge sorters can be constructed in a straightforward manner for any set of list sizes,
including \mbox{power-of-2} list sizes.

When merging more than 2 sorted input lists, it is perhaps less clear what the \mbox{state-of-the art} methodology is.
Here, the Multiway Merge Sort devices in \cite{us_access_multiway_2022}\cite{multiway_merge_patent}
are considered to be the \mbox{state-of-the-art} for merging more than 2 sorted lists.
The 3c\_7r 3-way LOMS device, discussed in Section~\ref{sec:list_offset_3_way_merge} and Appendix~\ref{app:3_way_merge_setup_design},
is evaluated relative to the comparable Multiway Merge Sort 3c\_7r device.

%% $$$$$$$$$$$$$$$$$$$$$$$$$$$$$$$$$$$$$$$$$$$$$$$$$$$$$$$$$$$$$$$$$$$$$$$$$$$$$

\subsection{2-way LOMS Designs}

\label{sec:list_offset_characterization_designs}

\begin{figure}[h]
  \centering
  \includegraphics[width=0.75\linewidth]{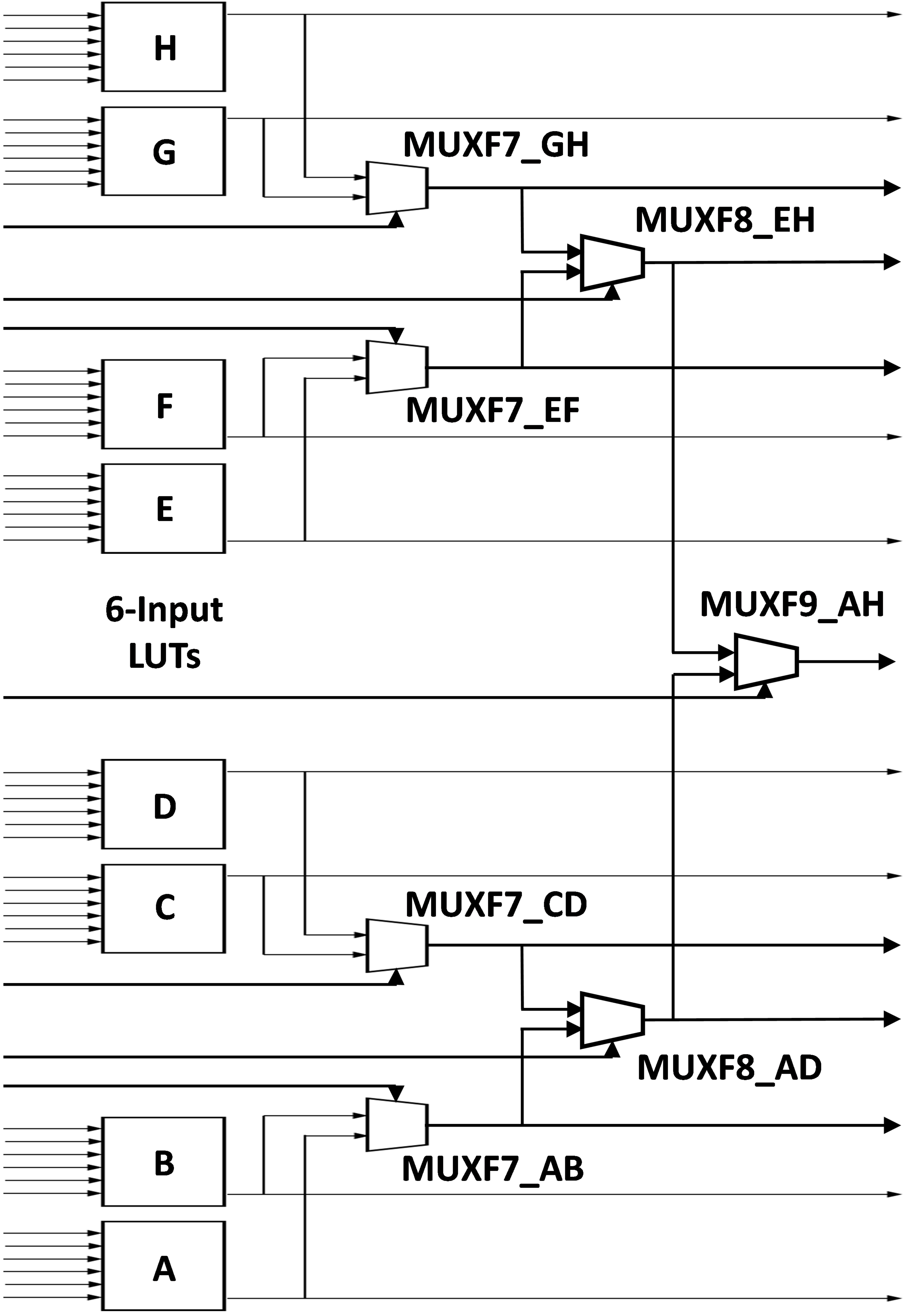}
  \caption{Ultrascale+ Slice: 8 LUTs + 3 levels of MUXF* multiplexers.}
  \label{fig:uscale_plus_8_LUT_slice}
\end{figure}

The merge sorters constructed for \mbox{2-way} merge sort characterization are designed to take advantage
of the MUXF* \mbox{2-to-1} multiplexers which are \mbox{hard-wired} in the Kintex Ultrascale+ slice,
as shown in Fig.~\ref{fig:uscale_plus_8_LUT_slice}.
It is assumed that use of these \mbox{hard-wired} MUXF* structures will produce faster devices,
versus those that send LUT outputs into the programmable FPGA fabric.
The S2SM and LOMS \mbox{2-way} merge sorter Verilog design files
specifically force use of the MUXF* structures in the Ultrascale+ device.

The Versal Prime device does not have the MUXF* structures in its slices,
but the AMD Vivado software translates MUXF* structures into general \mbox{2-to-1} multiplexers for Versal Prime,
so the same Verilog design files are still used.  The Versal Prime \mbox{2-to-1} multiplexers will be constructed
in a separate slice in series with the 2 LUTs providing the multiplexer inputs.

\mbox{Single-Stage} \mbox{2-way} Merge Devices are the only devices in an S2MS sorter,
but they are also used as the first stage column sorters for a \mbox{2-way} LOMS device.
Two different methodologies have been employed in this characterization
to produce S2MS sorters
which utilize the Fig.~\ref{fig:uscale_plus_8_LUT_slice} slice.

\begin{figure}[h]
  \centering
  \includegraphics[width=\linewidth]{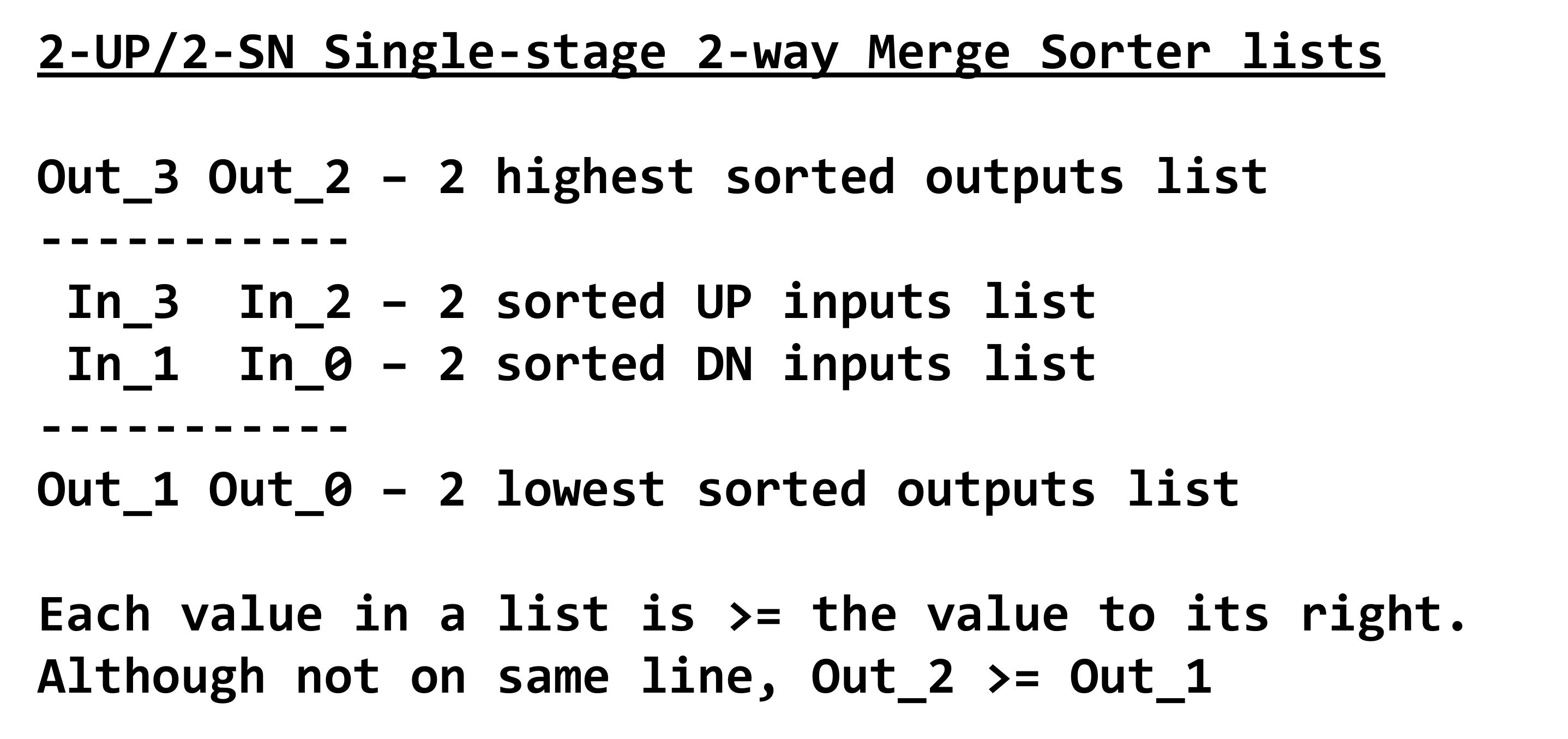}
  \caption{UP-2/DN-2 2-way Merge List Definitions and Order.}
  \label{fig:up_2_dn_2_list_defs}
\end{figure}

\begin{figure}[h]
  \centering
  \includegraphics[width=0.80\linewidth]{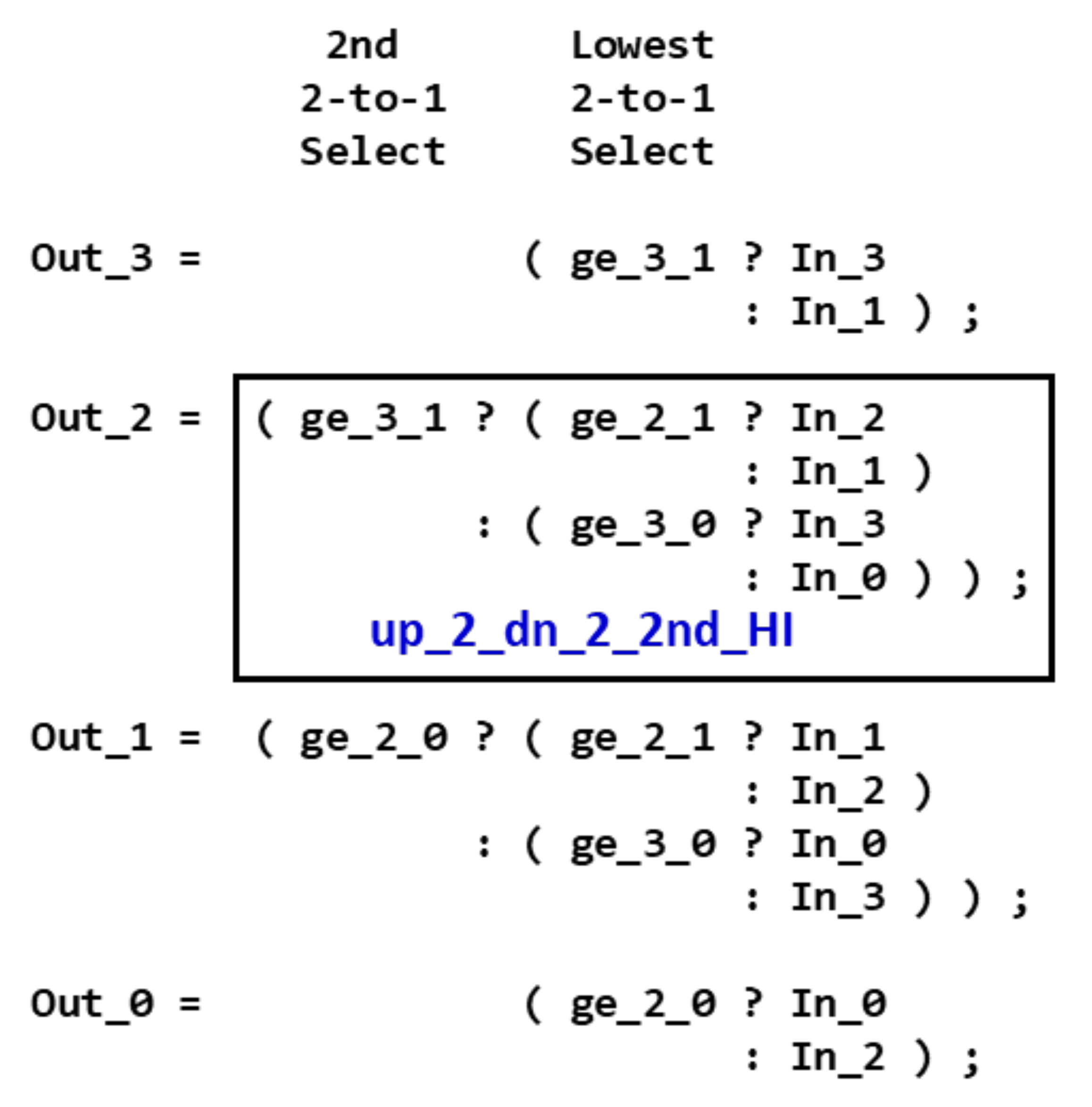}
  \caption{UP-2/DN-2 Output Equations using Conditional Operators.}
  \label{fig:up_2_dn_2_output_equations}
\end{figure}

The simplest S2MS sorter, called an \mbox{UP-2/DN-2} device, merges 2 lists each with 2 values,
and it will be used to illustrate the 2 methodologies.
Fig.~\ref{fig:up_2_dn_2_list_defs} shows the definitions of the input and output \mbox{UP-2/DN-2} lists, and
Fig.~\ref{fig:up_2_dn_2_output_equations} shows the equations used to define the 4 output values, using the principles
introduced in \cite{s2sm_Asilomar}\cite{s2sm_patent}.
All 4 inputs can go to the middle 2 outputs, Out\_2 and Out\_1, and 3 ge\_* comparison signals determine
which input goes to each of those outputs.  Since there are 7 signals in each output equation,
neither output equation can be simply implemented in a \mbox{6-input} LUT.

In the 2insLUT methodology, 2 inputs and 1 ge\_* comparison signal are inputs to a single LUT.
For Out\_2, the In\_2 and In\_1 input bits, plus comparison signal ge\_2\_1, are the 3 inputs to one LUT.
In\_3 and In\_0 input bits, plus comparison signal ge\_3\_0, are the 3 inputs to an adjacent LUT.
Comparison signal ge\_2\_0 is then used as the select line to the MUXF7 which combines the outputs of those 2 LUTs.

The second design methodology is called 4insLUT, where 4 input bits are input signals to the same LUT.
The other 2 inputs to the \mbox{6-input} LUT are one of the 3 comparison signal inputs,
plus another signal which is a function of the other 2 comparison signals.
For Out\_2, ge\_2\_0 is still a LUT input, but the last input is the function signal [ ge\_2\_1 || ( ! ge\_3\_0 ) ].
This function signal is produced in a LUT that is in series with the LUT containing
all 4 input bits and the other comparison signal as inputs.
Since this signal is produced in series with the \mbox{4-input} LUT, it produces a slower output,
while enabling a denser \mbox{4-input} bits per LUT design.

In short, the 2insLUT methodology tends to produce a fast \mbox{2-way} merge, with 3 inputs per LUT.
The 4insLUT method produces a slower merge sort, but with fewer resources needed
for a given size S2MS device. Both methodologies were characterized, and their speed and resource usage differences
are discussed in the results section.

\begin{figure}[h]
  \centering
  \includegraphics[width=\linewidth]{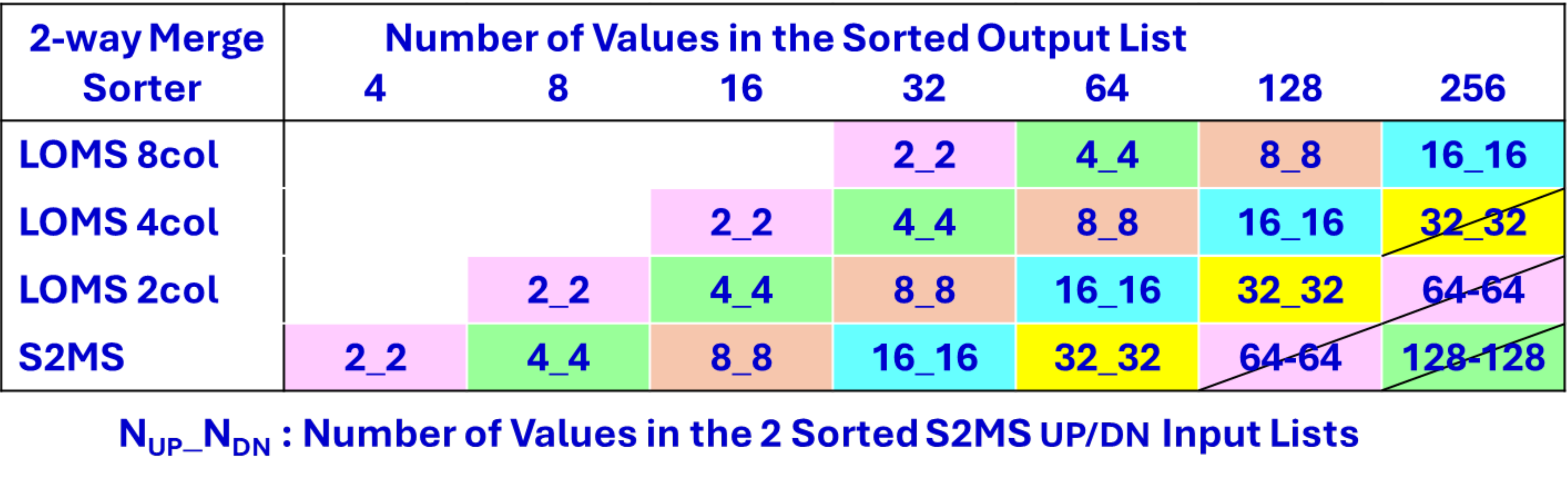}
  \caption{Matrix of S2MS devices in S2MS/LOMS Merge Sorters.}
  \label{fig:s2ms_matrix}
\end{figure}

Fig.~\ref{fig:s2ms_matrix} shows how the various \mbox{2-way} S2MS and LOMS merge sorters
were constructed for the characterization work here,
for devices with up to 256 outputs.
The N$_{UP}$\_N$_{DN}$ labels indicate the type of S2MS \mbox{2-way} merge sorter used in both S2MS and LOMS devices.
Several cells on the right and bottom table edge have a diagonal line running through the cell.
These marked cells indicate \mbox{2-way} merge
sorters that could not pass \mbox{place-and-route} for the \mbox{32-bit} Ultrascale+ 2insLUT merge sorters
whose results will be reported in Section~\ref{sec:loms_2ins_2_way_results}.  It should be clear from these marked cells,
and the data in Section~\ref{sec:loms_2ins_2_way_results}, that it is difficult
to implement devices that contain large S2MS \mbox{2-way} merge sorters.

%% $$$$$$$$$$$$$$$$$$$$$$$$$$$$$$$$$$$$$$$$$$$$$$$$$$$$$$$$$$$$$$$$$$$$$$$$$$$$$

\subsection{3-way LOMS Design}

\label{sec:list_offset_3_way_characterization_design}

Only one \mbox{k-way} List Offset Merge Sorter is characterized here, with k>2, and compared to the equivalent Multiway Merge Sorter.
This is the 3c\_7r merge sorter, which merges 3 input sorted lists, each with 7 values, which was the subject of
Section~\ref{sec:list_offset_3_way_merge}.  Both the full merge of all 21 values, and the faster median merge, are characterized here
for both \mbox{8-bit} and \mbox{32-bit} values, targeting both FPGA devices.

%% $$$$$$$$$$$$$$$$$$$$$$$$$$$$$$$$$$$$$$$$$$$$$$$$$$$$$$$$$$$$$$$$$$$$$$$$$$$$$

\section{Results}

\label{sec:top_results}

Various merge sorters have been constructed and synthesized for both \mbox{8-bit} and \mbox{\mbox{32-bit}} unsigned integers
in 2 FPGA products, from 2 distinctly different FPGA families.  Combinatorial propagation delay (speed) and LUT resource usage data
have been gathered for these merge sorters, and the results are presented and discussed in the sections below.

For \mbox{2-way} merge sort, results have been collected and compared for \mbox{2-way} List Offset Merge Sorters,
\mbox{Single-Stage} \mbox{2-way} Merge Sorters, and Kenneth Batcher's Bitonic Merge Sorters and \mbox{Odd-Even} Merge Sorters.
The output list sizes are a \mbox{power-of-2}, and the input list
sizes are equal, half the size of the output lists, and clearly a \mbox{power-of-2} as well.
Batcher's 2 types of merge sorters are easiest to define, and most efficient, when the list sizes are a \mbox{power-of-2}.
The number of output values are plotted on the \mbox{x-axis} for all \mbox{2-way} merge figures,
and a logarithmic scale is used for these \mbox{x-axes}.

For \mbox{k-way} merge sort, with $k{>}2$, one List Offset \mbox{3-way} Merge Sorter has been characterized relative to
the comparable \mbox{state-of-the-art} Multiway Merge Sorter.  The LOMS \mbox{\mbox{3-way}} merge sorter used for this analysis is the 3c\_7r
merge sorter defined in Section~\ref{sec:list_offset_3_way_merge}.

FPGA data for the various merge sorters has been generated for the \mbox{xcku5p-ffva676-3-e} from the AMD Kintex Ultrascale+ family
and the AMD \mbox{xcvm1102-sfva784-2HP-i-S} from the Versal Prime family.
The synthesis results were obtained using the AMD 2024.2 Vivado tool.  Neither product needs a purchased
license to run on 2014.2, so designers and researchers should have no trouble verifying the results presented here.

As mentioned above, speed results plots use combinatorial propagation delay, the time required for the slowest signal to propagate
from a sorter input port to an output port.  The fastest designs have the lowest propagation delay,
and the slowest designs have the highest propagation delay.  Propagation delay is plotted using a linear \mbox{y-axis}.
LUT resource usage plots use a logarithmic \mbox{y-axis}.

The recently introduced \mbox{Single-Stage} \mbox{2-way} Merge Sorters were previously compared to Batcher's \mbox{2-way} merge sorters,
using \mbox{8-bit} data generated for a single Ultrascale+ product.
Section~\ref{sec:s2ms_vs_Batcher_2_way_results} below presents data and analysis
for these existing \mbox{2-way} Merge Sorters, now constructed in the 2 distinctly different FPGAs,
using both \mbox{8-bit} and \mbox{32-bit} values.

Data for the \mbox{2-way} List Offset Merge Sorters introduced here is initially presented in Section~\ref{sec:loms_4ins_2_way_results}.
The data in this section utilizes the 4insLUT methodology, and focuses on small S2SM and LOMS merge sorters, which are noticeably
faster than comparable Bitonic Merge Sorters, and even use fewer resources.

The data for the \mbox{2-way} List Offset Merge Sorters
covered in Section~\ref{sec:loms_4ins_2_way_results} uses the 2insLUT methodology, and focuses on larger \mbox{2-way} Merge Devices.
The S2SM devices are faster than comparable LOMS devices, but use a large number of resources.
The 2insLUT \mbox{2-way} LOMS devices are significantly faster than Batcher merge sorters,
and use fewer resources than comparable S2SM devices.
In fact, some proposed larger S2SM \mbox{2-way} merge devices will not fit in a targeted FPGA, but the same size LOMS devices will fit.

The 3c\_7r \mbox{3-way} merge defined in Section~\ref{sec:list_offset_3_way_merge} are also constructed and synthesized in the 2 target FPGAs,
and their results are compared to the matching \mbox{state-of-the-art} Multiway Merge Sorting Network \mbox{3-way} devices.
These \mbox{3-way} results are presented in Section~\ref{sec:3_way_results}.

%% $$$$$$$$$$$$$$$$$$$$$$$$$$$$$$$$$$$$$$$$$$$$$$$$$$$$$$$$$$$$$$$$$$$$$$$$$$$$$

\subsection{S2MS vs Batcher 8-bit/32-bit Merge on 2 FPGAs}

\label{sec:s2ms_vs_Batcher_2_way_results}

In the recent publication introducing \mbox{Single-Stage} \mbox{2-way} Merge Sort \cite{s2sm_Asilomar},
the new S2SM merge sorters were compared to Batcher's \mbox{state-of-the-art} \mbox{2-way} merge sorters,
using designs constructed for \mbox{8-bit} unsigned values and synthesized for one Ultrascale+ product.

\begin{figure}[h]
  \centering
  \includegraphics[width=\linewidth]{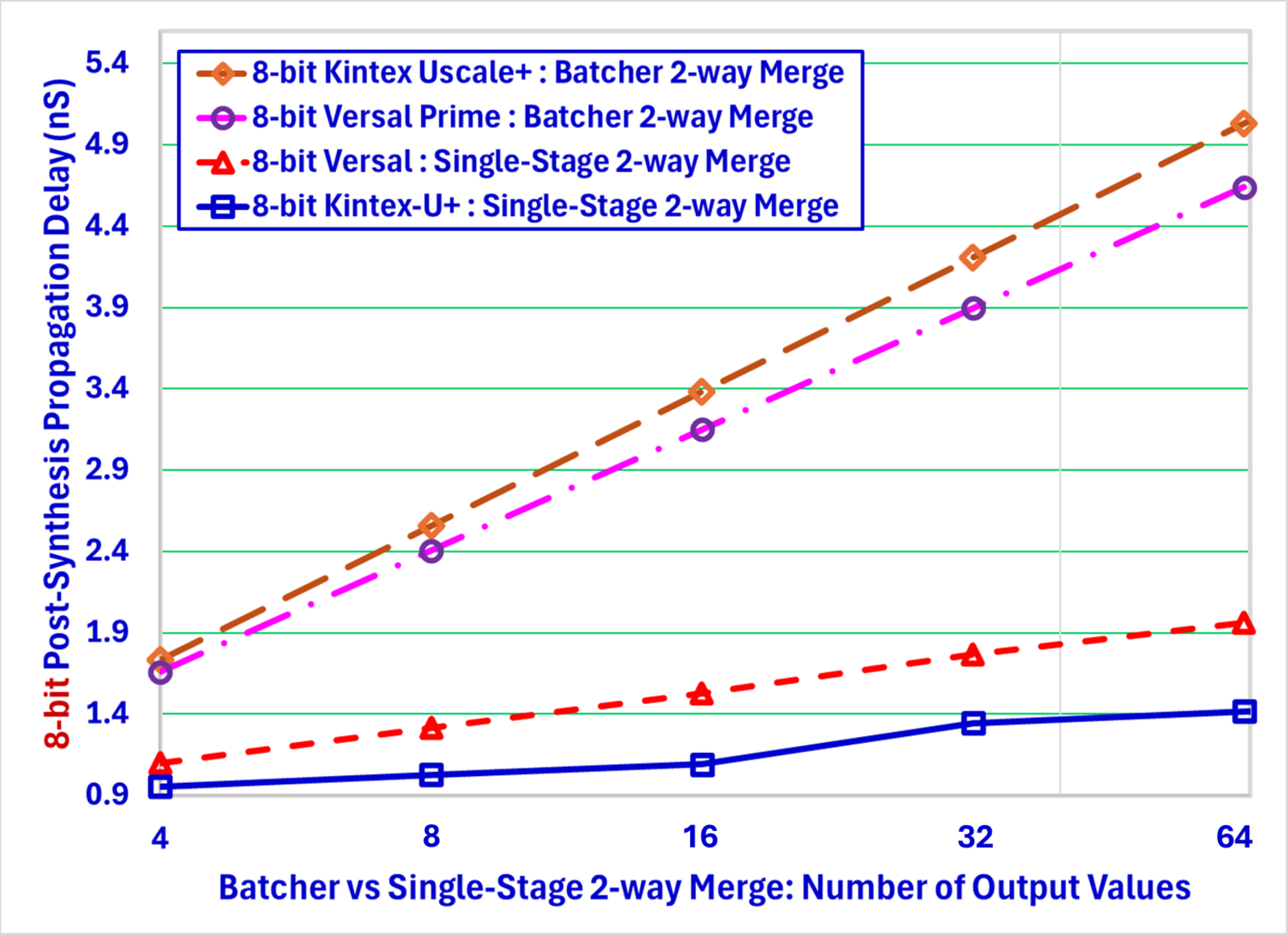}
  \caption{Batcher vs Single-Stage 2-way Merge Speed: 8-bit Values.}
  \label{fig:old_comps_8_bit_speed}
\end{figure}

In this work, data has been gathered for both \mbox{8-bit} and \mbox{32-bit} values, using S2SM and Batcher \mbox{2-way} merge sorters,
now constructed in 2 distinctly different FPGA products.
Fig.~\ref{fig:old_comps_8_bit_speed} displays Batcher vs S2MS \mbox{8-bit} speed data,
for both the Kintex Ultrascale+ and the Versal Prime devices.  The top 2 curves identify the slowest devices, as they have the highest
propagation delay.  Batcher's \mbox{Odd-Even} Merge Sort and Bitonic Merge Sort devices
have identical propagation delay results for a given FPGA,
so their identical values are plotted as ``Batcher'' results.

The bottom 2 curves separately identify \mbox{Single-Stage} \mbox{2-way} Merge speed results
for Versal Prime and Kintex Ultrascale+ devices.
The two devices use the same Verilog design files, but the MUXF* structures,
which combine 2 internal slice signals into a 3rd internal slice signal, only exist for the Ultrascale+ device.
For the Versal Prime merge sorters, the 2 inputs to a MUXF* structure are sent to the programmable
interconnect and then combined in a LUT in series with the LUTs providing the 2 MUXF* inputs.  For the Versal implementation,
this produces a slower path than the comparable Ultrascale+ path using the \mbox{hard-wired} MUXF* structure.

\begin{figure}[h]
  \centering
  \includegraphics[width=\linewidth]{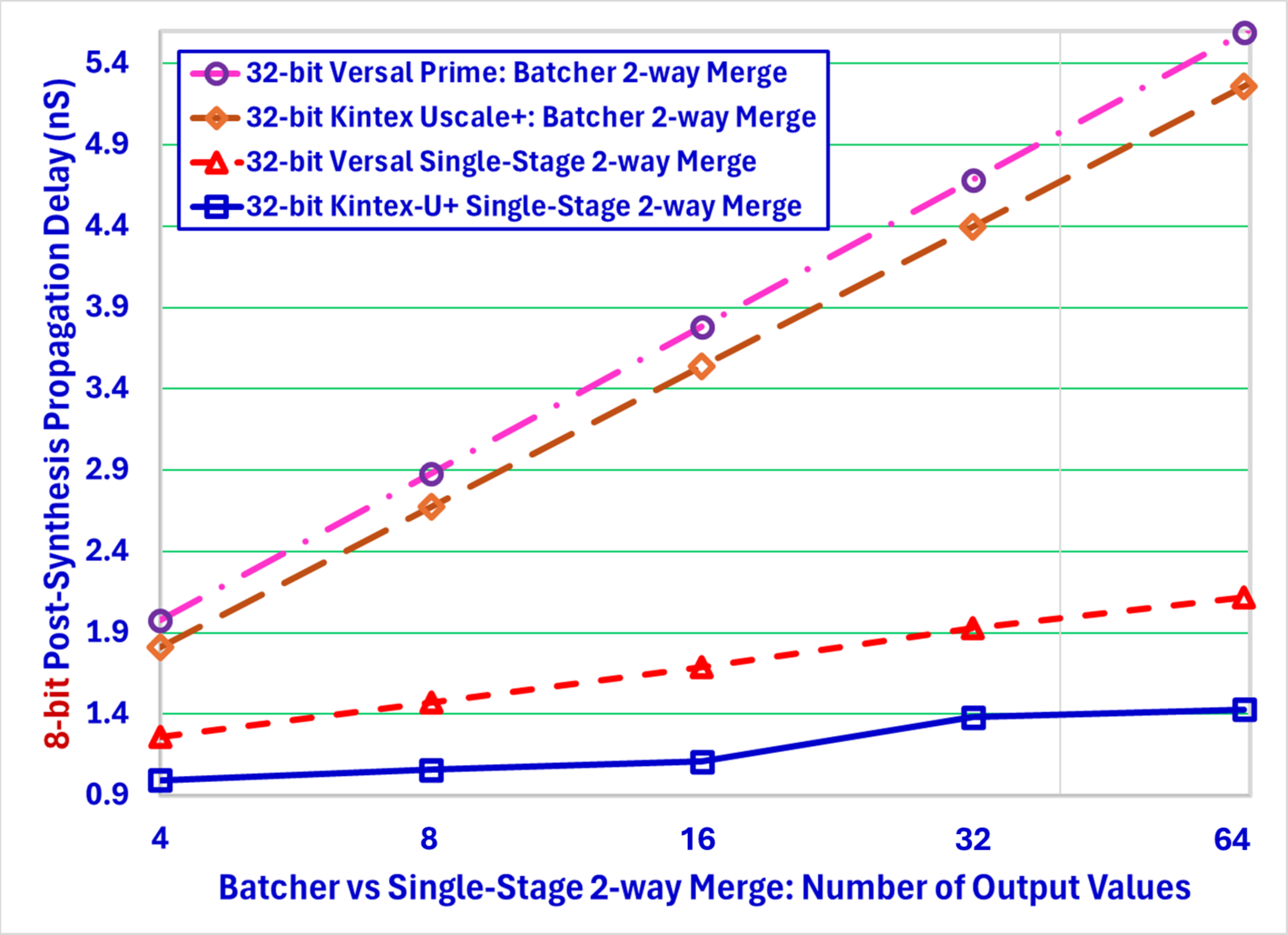}
  \caption{Batcher vs Single-Stage 2-way Merge Speed: 32-bit Values.}
  \label{fig:old_comps_32_bit_speed}
\end{figure}

\begin{figure}[h]
  \centering
  \includegraphics[width=\linewidth]{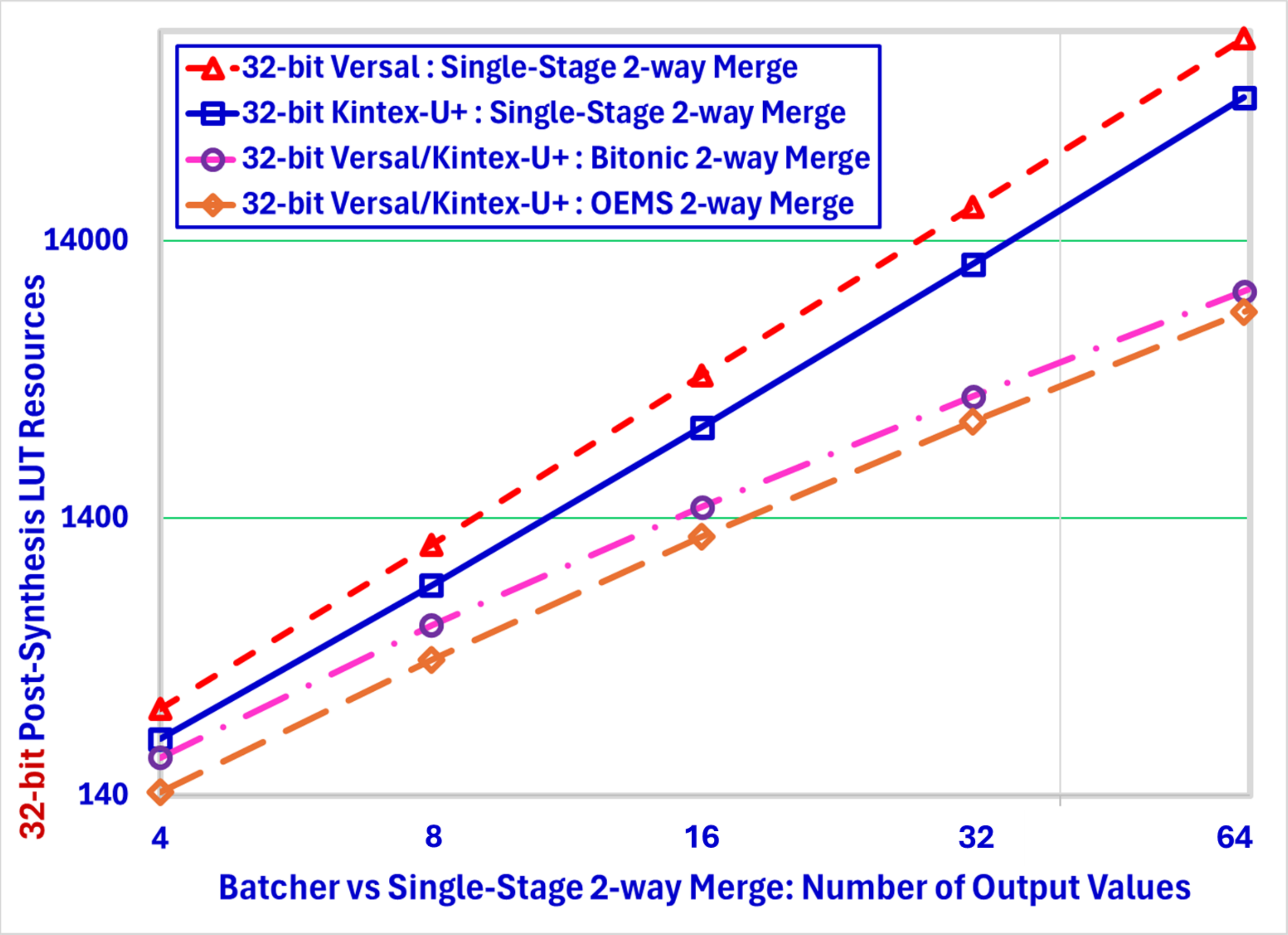}
  \caption{Batcher vs Single-Stage 2-way Merge LUTs: 32-bit Values.}
  \label{fig:old_comps_32_bit_LUTs}
\end{figure}

The lowest and therefore fastest Fig.~\ref{fig:old_comps_8_bit_speed} curve is
for \mbox{Single-Stage} \mbox{2-way} Merge Sorters constructed in the Ultrascale+ device with the 2insLUT design methodology.
This curve is fairly flat up to 16 output values, before it steps up
to another fairly flat section between 32 and 64 outputs.
Only 1 series slice is required to construct output multiplexers for up to 16 outputs, after the
comparison signals have been constructed in parallel.  For 32 and 64 outputs, however, 2 series slices are required to construct
the largest output multiplexers.
The path in the programmable interconnect between the 1st and 2nd
series slices creates the step up in this curve.

The Fig.~\ref{fig:old_comps_8_bit_speed} Versal \mbox{Single-Stage} \mbox{2-way} Merge Sort curve has a consistent slope.
These devices use the same Verilog design files as the Ultrascale+ device,
but the Versal slice
does not have the MUXF* structures to combine LUT outputs internally.
Each LUT output in a \mbox{multi-slice} path must exit its
slice and propagate
through the programmable interconnect to the next series slice.
Each doubling of the number of output values requires an \mbox{additional Versal} series slice for the slowest paths.

Fig.~\ref{fig:old_comps_32_bit_speed} is the same type of plot as Fig.~\ref{fig:old_comps_8_bit_speed},
but the Fig.~\ref{fig:old_comps_32_bit_speed} data results are for \mbox{32-bit} data.
The top 2 Fig.~\ref{fig:old_comps_32_bit_speed} curves may appear to match the top 2
Fig.~\ref{fig:old_comps_8_bit_speed} curves, but there is a significant difference.
For the \mbox{32-bit} data, the Versal Prime Batcher speed is slower than the Ultrascale+ speed, which is the reverse
of what was seen for the \mbox{8-bit} data in Fig.~\ref{fig:old_comps_8_bit_speed}.

The bottom 2 Fig.~\ref{fig:old_comps_32_bit_speed} curves also show that the
Versal \mbox{32-bit} S2SM delay is noticeably higher (and therefore slower)
relative to its location for the Fig.~\ref{fig:old_comps_8_bit_speed} \mbox{8-bit} data.
As was seen with the Batcher curves, the Ultrascale+ S2SM Fig.~\ref{fig:old_comps_32_bit_speed} \mbox{32-bit} curve
changed very little relative to its \mbox{8-bit} curve in Fig.~\ref{fig:old_comps_8_bit_speed} .

Fig.~\ref{fig:old_comps_32_bit_LUTs} is a \mbox{log-log} plot of the \mbox{32-bit} resources used by the 
Fig~\ref{fig:old_comps_32_bit_speed} \mbox{2-way} merge devices.  A comparable plot is not shown for \mbox{8-bit} values,
as it would look much like Fig.~\ref{fig:old_comps_32_bit_LUTs}.
The Fig.~\ref{fig:old_comps_32_bit_LUTs} Batcher curves are separated into an OEMS and a Bitonic curve.  The number of LUT resources used for each type of Batcher merge sorter is identical when implemented in the Ultrascale+ and Versal FPGAs.
The 2 Batcher merge sorters use the fewest resources overall.

The Ultrascale+ S2SM devices use fewer resources than the Versal S2SM devices,
even though the Ultrascale+ S2SM devices are the overall fastest \mbox{2-way} merge devices.
This is due to the fact that Ultrascale+ LUT outputs can route \mbox{hard-wired} internally in a slice to MUXF* structures,
while comparable Versal LUT outputs must route to additional LUTs in a series slice, using the programmable interconnect.

%% $$$$$$$$$$$$$$$$$$$$$$$$$$$$$$$$$$$$$$$$$$$$$$$$$$$$$$$$$$$$$$$$$$$$$$$$$$$$$

\subsection{List Offset 4insLUT 2-way Merge Results}

\label{sec:loms_4ins_2_way_results}

\begin{figure}[h]
  \centering
  \includegraphics[width=\linewidth]{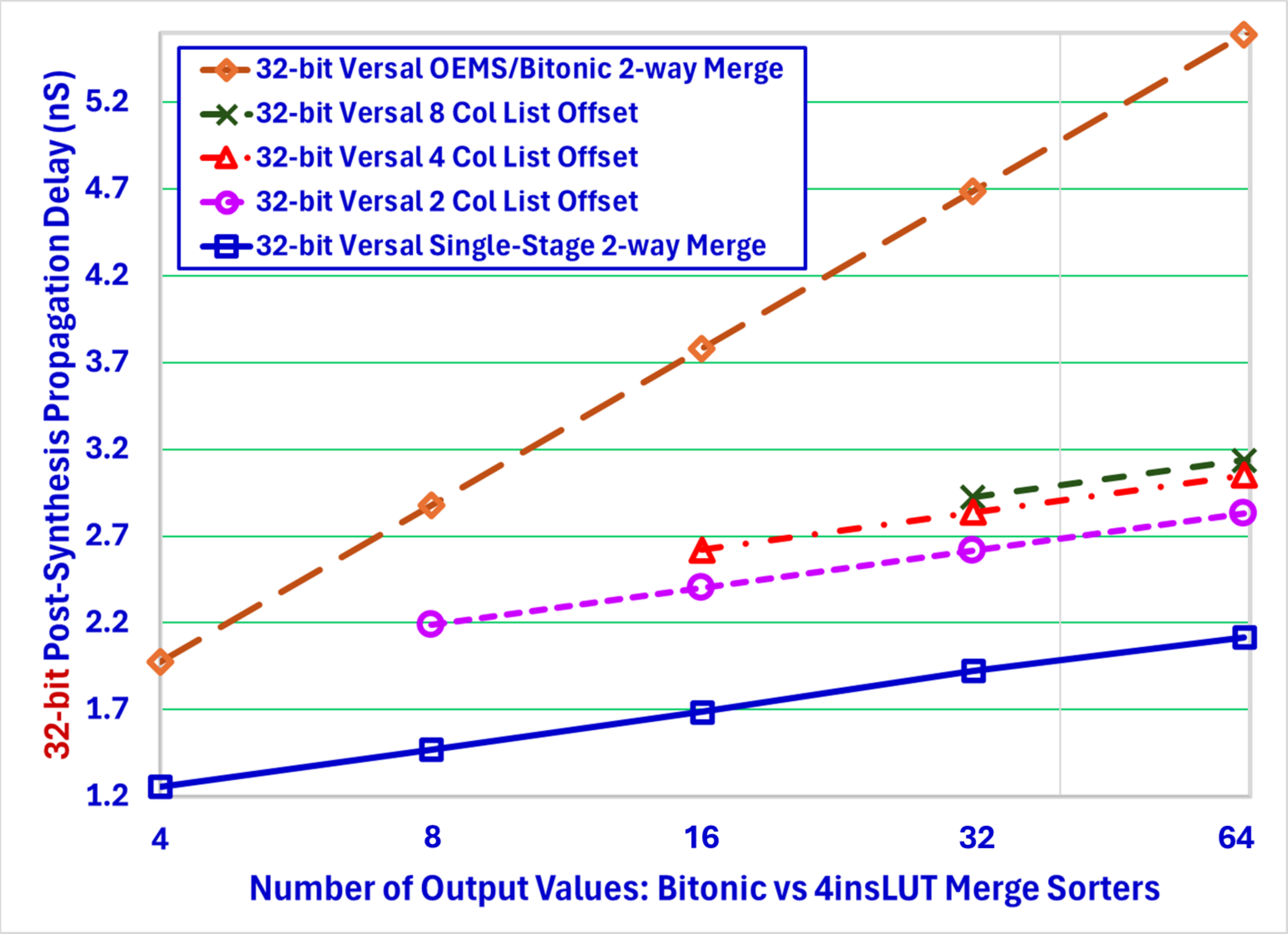}
  \caption{Bitonic vs 4insLUT S2SM and LOMS Speed: 32-bit Values.}
  \label{fig:LUT4ins_Versal_32_bit_speed}
\end{figure}

\begin{figure}[h]
  \centering
  \includegraphics[width=\linewidth]{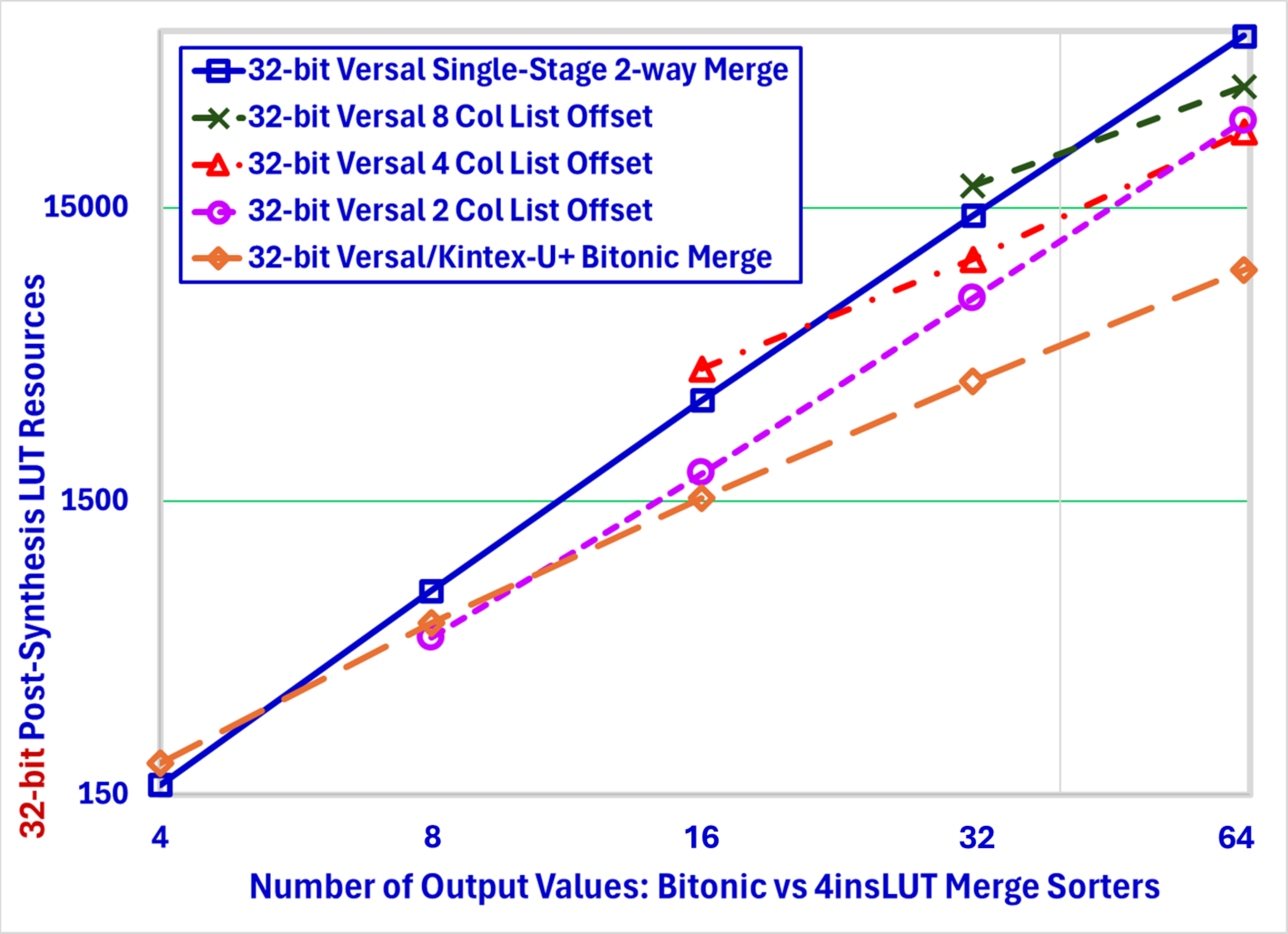}
  \caption{32-bit Bitonic vs 4insLUT S2SM and LOMS LUT Resources.}
  \label{fig:LUT4ins_Versal_32_bit_LUTs}
\end{figure}

Figs.~\ref{fig:LUT4ins_Versal_32_bit_speed}~and~\ref{fig:LUT4ins_Versal_32_bit_LUTs}
display the propagation delay and LUT resource results for Bitonic Merge Sort as well as 4insLUT
methodology S2SM and \mbox{2-way} LOMS merge sorters, which were discussed in
Section~\ref{sec:list_offset_characterization_designs}.
The data in these figures were produced from \mbox{32-bit} Versal merge sorters.

The 4insLUT merge sorters, which tend to pack 4 input bits into a LUT,
use fewer resources than 2insLUT sorters, which separate the 4 input bits into 2 LUTs.
As shown in Fig.~\ref{fig:LUT4ins_Versal_32_bit_LUTs},
the S2SM \mbox{4-output} merge sorter uses fewer resources than the comparable Bitonic merge sorter,
even though Fig.~\ref{fig:LUT4ins_Versal_32_bit_speed} shows that it is significantly faster.
The \mbox{2-column} LOMS \mbox{8-output} merge sorter also uses fewer resources than the comparable Bitonic merge sorter,
even though it, too, is significantly faster.
In addition, the fast \mbox{2-column} LOMS \mbox{16-output} merge sorter uses only slightly
more resources than the comparable Bitonic merge sorter.

It can be argued that, for a large Bitonic merge sorter, a Bitonic \mbox{4-output} merge sorter should be replaced
by the matching 4insLUT S2SM device, and the \mbox{8-output}, if not the \mbox{16-output}, Bitonic merge sorters should be
replaced by the comparable \mbox{2-column} 4insLUT LOMS merge sorters.
Using the \mbox{4insLUT} methodology for larger merge sorters is not emphasized in this work,
as they are slower than 2insLUT merge sorters,
and the larger 4insLUT merge sorters can have routing congestion problems,
while comparable 2insLUT merge sorters tend not to have routing congestion.

%% $$$$$$$$$$$$$$$$$$$$$$$$$$$$$$$$$$$$$$$$$$$$$$$$$$$$$$$$$$$$$$$$$$$$$$$$$$$$$

\subsection{List Offset 2insLUT 2-way Merge Results}

\label{sec:loms_2ins_2_way_results}

\begin{figure}[h]
  \centering
  \includegraphics[width=\linewidth]{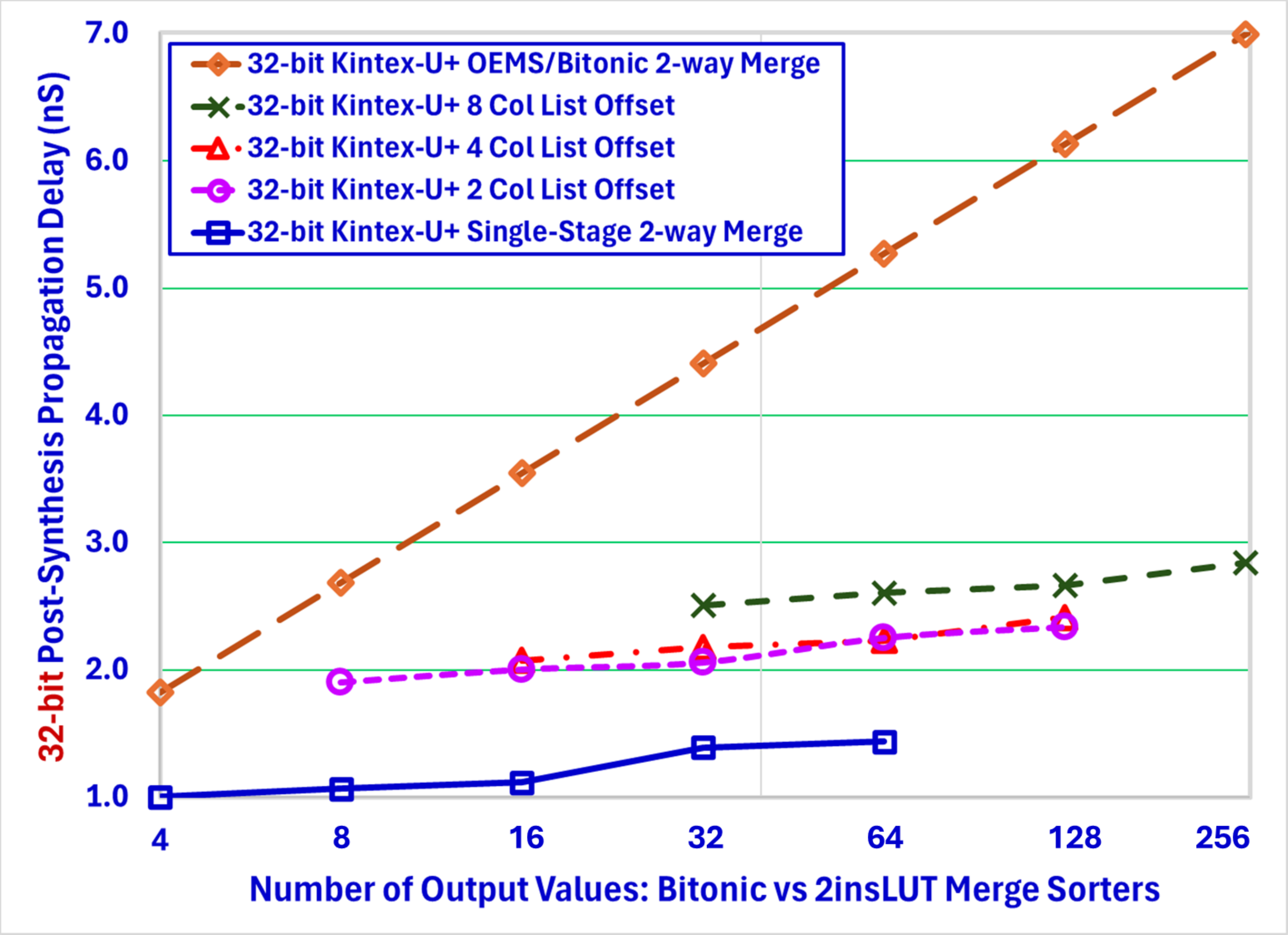}
  \caption{Bitonic vs 2insLUT S2SM and LOMS Speed: 32-bit Values.}
  \label{fig:LUT2ins_KinUp_32_bit_speed}
\end{figure}

\begin{figure}[h]
  \centering
  \includegraphics[width=\linewidth]{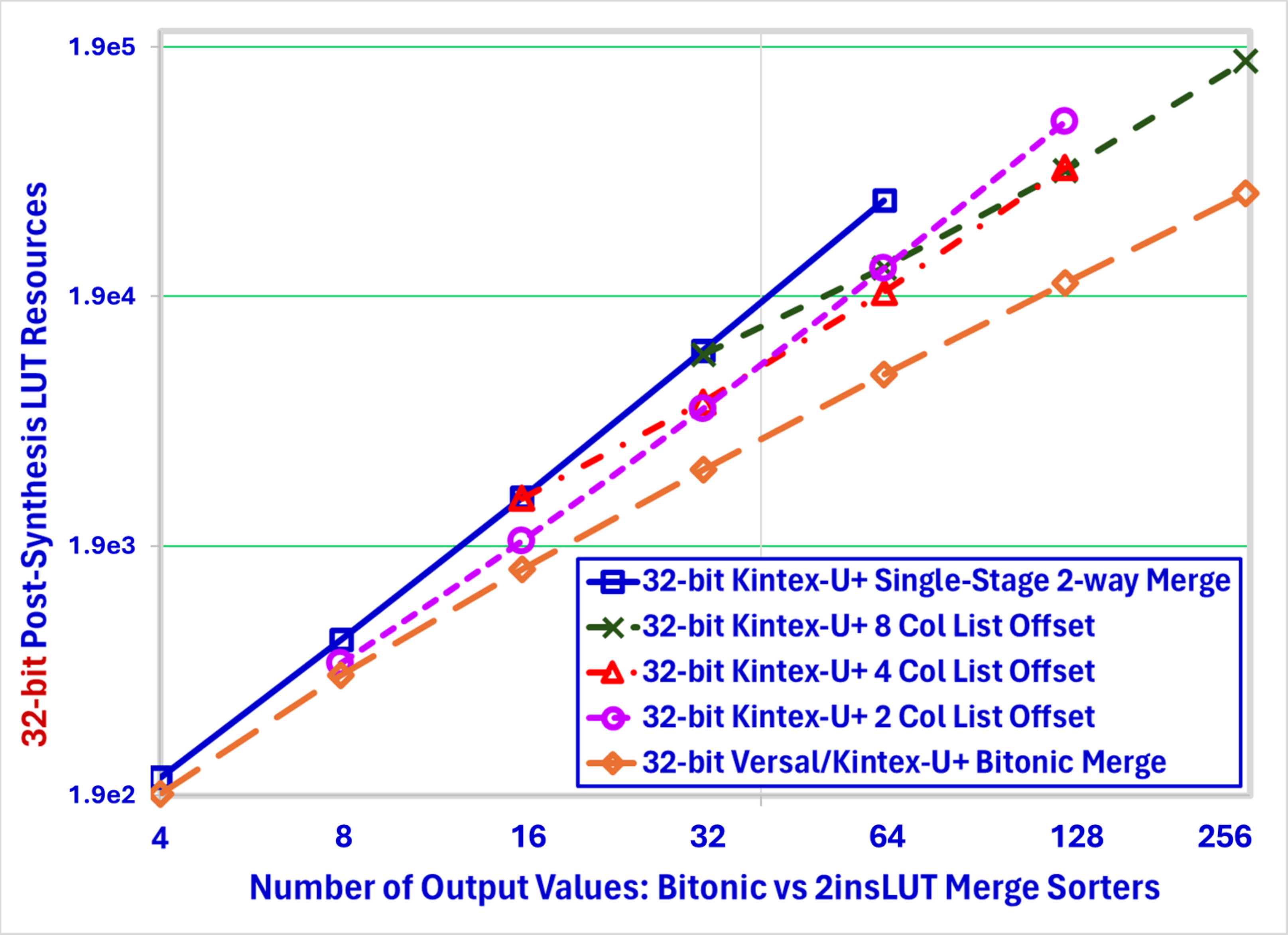}
  \caption{32-bit Bitonic vs 2insLUT S2SM and LOMS LUT Resources.}
  \label{fig:LUT2ins_KinUp_32_bit_LUTs}
\end{figure}

Figs.~\ref{fig:LUT2ins_KinUp_32_bit_speed}~and~\ref{fig:LUT2ins_KinUp_32_bit_LUTs} display \mbox{32-bit} \mbox{Kintex-Ultrascale+}
propagation delay and LUT resource results for S2SM and \mbox{2-way} LOMS merge sorters which utilize 2insLUT methodology,
as described in Section~\ref{sec:list_offset_characterization_designs}, as well as comparable Bitonic Merge Sorters.
The 2insLUT \mbox{Kintex-Ultrascale+} merge sorters are designed to take advantage of the MUXF* multiplexer structures
\mbox{hard-wired} in an Ultrascale+ slice.

The 2insLUT Figs.~\ref{fig:LUT2ins_KinUp_32_bit_speed}~and~\ref{fig:LUT2ins_KinUp_32_bit_LUTs} focus on larger merge sorters than
were displayed in Figs.~\ref{fig:LUT4ins_Versal_32_bit_speed}~and~\ref{fig:LUT4ins_Versal_32_bit_LUTs}, so the logarithmic \mbox{x-axis}
extends to 256 outputs for these 2 figures.  The output multiplexers of the Fig.~\ref{fig:LUT2ins_KinUp_32_bit_speed} S2SM sorters,
and the \mbox{Single-Stage} Merge Sort column sorters in the LOMS devices, require only 1 to 2 series slices,
so these curves are flatter than the comparable Versal LUT4ins curves in Fig.~\ref{fig:LUT4ins_Versal_32_bit_speed}.

The S2SM merge sorters are clearly the fastest, but also tend to use the most LUT resources.  The 64-output \mbox{single-stage} S2SM
was the largest merge sorter that could be fit into this particular Ultrascale+ FPGA, but 128-output 2col and 4col and 256-output
8col List Offset merge sorters were able to be constructed.  These larger \mbox{2-stage} List Offset merge sorters are significantly
faster than the Batcher merge sorters, so they are choices for speed, when S2SM merge sorters are unavailable.
For smaller output list sizes when S2SM devices are available, 2insLUT LOMS devices which meet speed targets may still
be the merge sorters of choice, due to their reduced LUT usage.

%% $$$$$$$$$$$$$$$$$$$$$$$$$$$$$$$$$$$$$$$$$$$$$$$$$$$$$$$$$$$$$$$$$$$$$$$$$$$$$

\subsection{3-way Merge Results}

\label{sec:3_way_results}

\begin{figure}[h]
  \centering
  \includegraphics[width=0.9\linewidth]{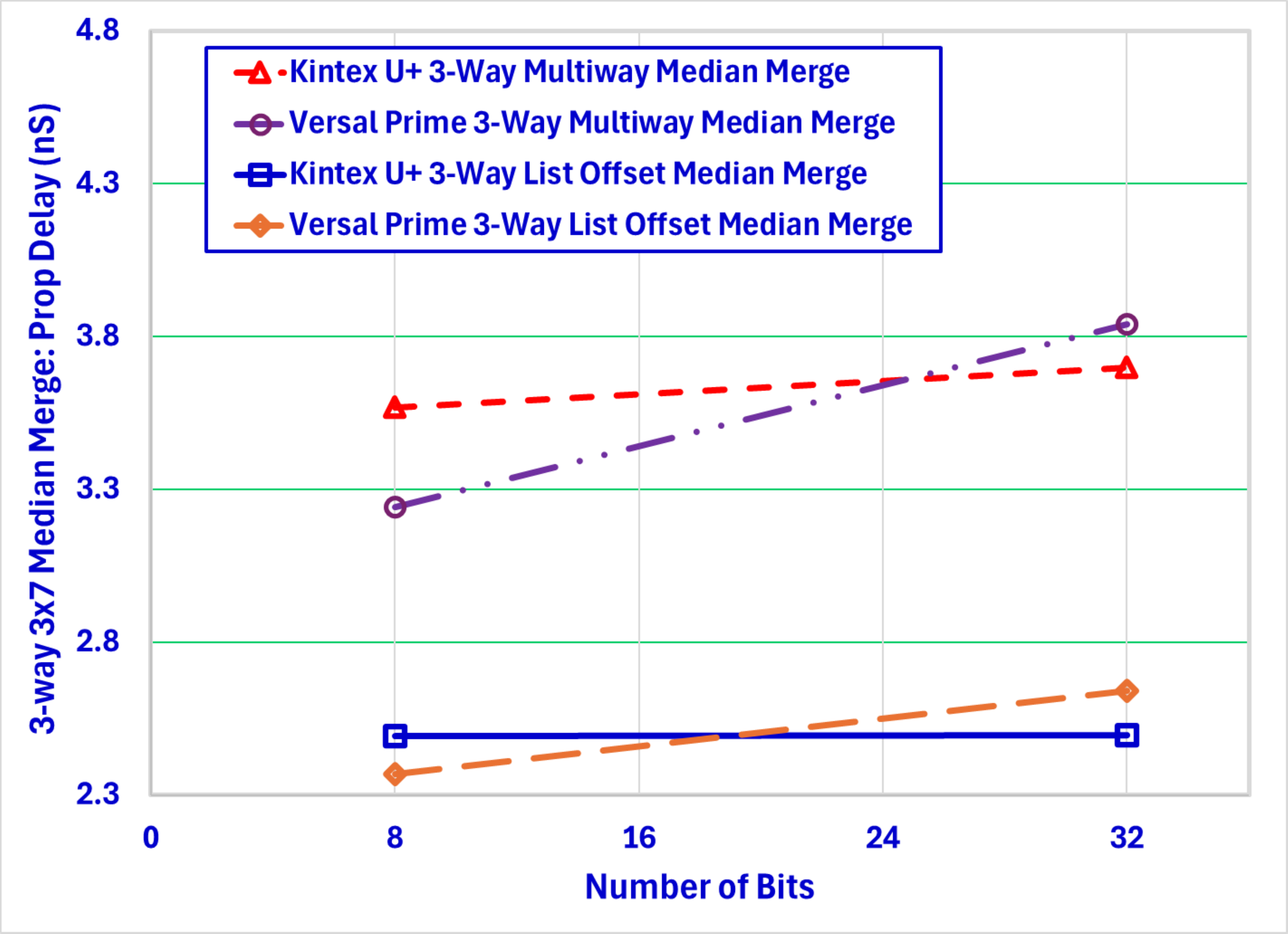}
  \caption{3c\_7r 3-way Median Merge Propagation Delays in nS.}
  \label{fig:3c_7r_median_plots}
\end{figure}

\begin{figure}[h]
  \centering
  \includegraphics[width=0.9\linewidth]{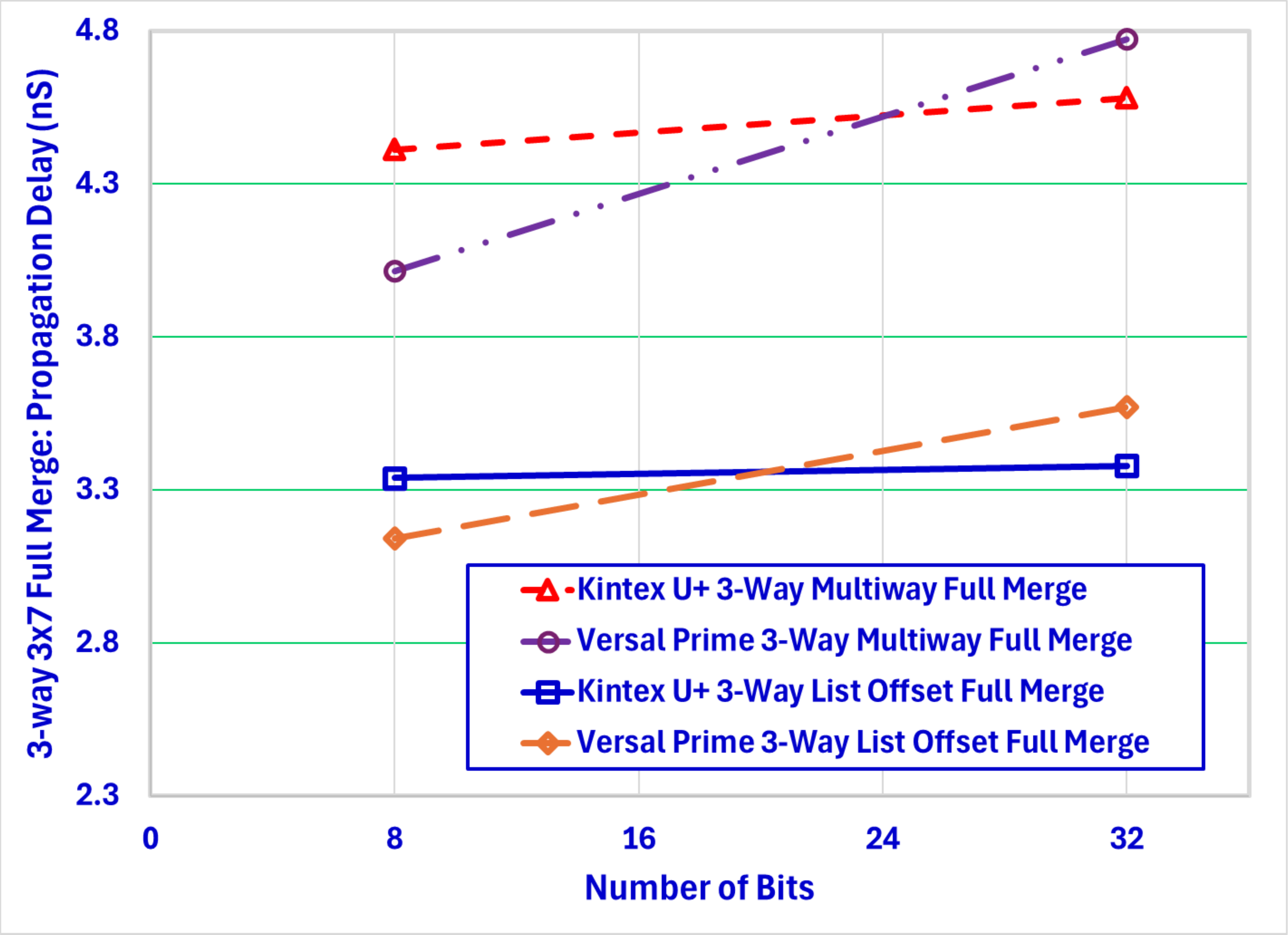}
  \caption{3c\_7r 3-way Full Merge Propagation Delays in nS.}
  \label{fig:3c_7r_full_plots}
\end{figure}

The propagation delay results for the 3c\_7r \mbox{3-way} merge sorters are shown
in Fig.~\ref{fig:3c_7r_median_plots} for median filter merge
and Fig.~\ref{fig:3c_7r_full_plots} for a full merge sort of 21 values.  Both figures have the same x and y axes.
All of the \mbox{3-way} merge figures have linear x and linear y axes.

In both figures, for both List Offset and Multiway Merge Sort devices,
Versal Prime devices are the fastest for \mbox{8-bit} devices, but are significantly slower for \mbox{32-bit} devices.
Kintex Ultrascale+ \mbox{32-bit} devices are only slightly slower than \mbox{8-bit} devices,
and the Kintex Ultrascale+ \mbox{32-bit} devices are faster than the Versal Prime \mbox{32-bit} devices.

\begin{figure}[h]
  \centering
  \includegraphics[width=0.9\linewidth]{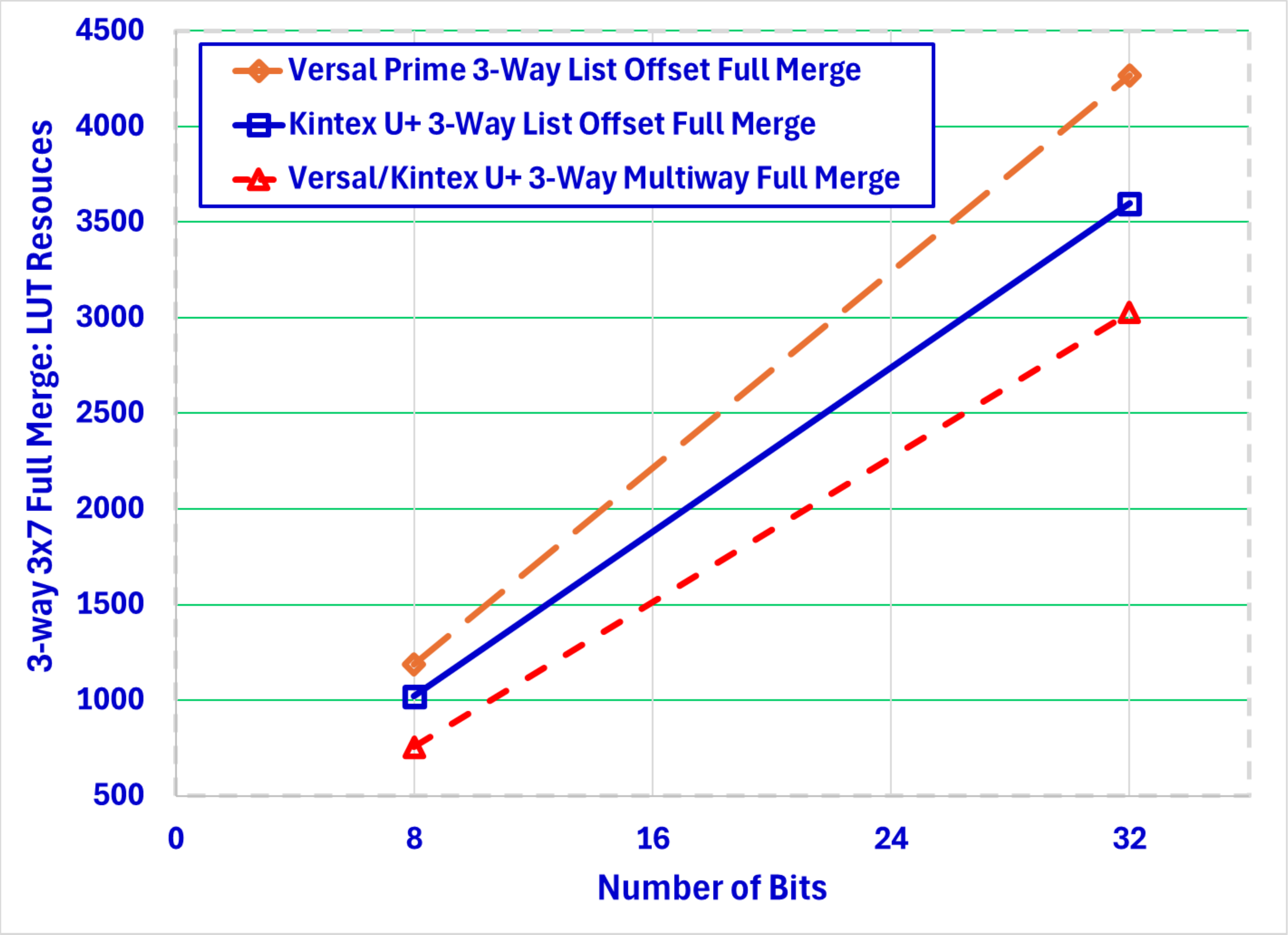}
  \caption{3c\_7r 3-way Full Merge LUT Resources.}
  \label{fig:3c_7r_full_LUTs}
\end{figure}

Determining the median requires only 2 stages for the List Offset devices,
and 4 stages for the \mbox{state-of-the-art} Multiway Merge Sorting Network devices.
The 2 Fig.~\ref{fig:3c_7r_median_plots} List Offset curves are the fastest, having the lowest propagation delays.
For these \mbox{median-determination} devices,
the \mbox{32-bit} speedups of List Offset versus comparable Multiway Merge Sort devices range from 1.45 to 1.48.

Fig.~\ref{fig:3c_7r_full_plots} curves are similar to those Fig.~\ref{fig:3c_7r_median_plots},
except that the propagation delay curves are higher.
The Fig.~\ref{fig:3c_7r_full_plots} curves are for devices that fully sort the 21 values, which requires 3 stages for List Offset
and 5 stages for Multiway Merge Sorting Network devices.
Once again, the 2 List Offset curves are the fastest, having the lowest propagation delays.
For these full sort devices, the \mbox{32-bit} speedups of List Offset versus
comparable Multiway Merge Sort devices range from 1.34 to 1.36.

The LUT resource usage for the full \mbox{3-way} merge sorters is shown in Fig.~\ref{fig:3c_7r_full_LUTs}.
The Multiway Merge sorters use the same number of LUTs in the 2 FPGAs, so there is only one curve for the 2 products.
The Multiway Merge sorters do use fewer resources than the List Offset merge sorters.

No LUT resource usage figure is shown for \mbox{3-way} median merge.  It would look much like the full merge figure, except
that the median sorters use fewer LUTs.

%% $$$$$$$$$$$$$$$$$$$$$$$$$$$$$$$$$$$$$$$$$$$$$$$$$$$$$$$$$$$$$$$$$$$$$$$$$$$$$

\section{Conclusion}

\label{sec:conclusion}

A new design methodology, List Offset Merge Sort, has been introduced in order to design, implement, and operate devices which can
improve merging 2 or more sorted input lists into a single sorted output list.  For \mbox{2-way} merge of sorted input lists, the new
LOMS devices are faster than Batcher's \mbox{2-way} merge, and some small LOMS merge sorters use fewer or similar LUT resources than
matching Bitonic merge sorters.  In addition, larger \mbox{2-way} LOMS devices, faster than Batcher devices,
can be implemented in an FPGA when matching \mbox{Single-Stage} Merge Sorters are too large to be fit into the FPGA.

LOMS devices can also merge more than 2 sorted input lists.  A \mbox{3-way} LOMS device is shown to fully sort 3 input lists more
quickly than the \mbox{state-of-the-art}, and to determine the median of input list values even more quickly.

Merge sorters designs using \mbox{Single-Stage} \mbox{2-way} Merge,
which take advantage of \mbox{hard-wired} structures in the Ultrascale+ slice,
were shown to be faster, while using fewer resources, than those implemented in a Versal Prime device, which does not have the
\mbox{hard-wired} structures in its slice.  Even in the Versal Prime product, however,
merge sorters using \mbox{Single-Stage} \mbox{2-way} Merge were shown
to be significantly faster than Batcher's \mbox{2-way} merge sorters, normally considered to be the \mbox{state-of-the-art}.

Both S2SM and LOMS are also much more versatile than Batcher's \mbox{2-way} merge devices.  Batcher's devices are difficult to design
unless both input lists have same number of values, and the number of values is a \mbox{power-of-2}.  There are no such restrictions
for both LOMS and S2SM devices.  They do not require that the input lists be the same size,
or that both are even or both are odd.

Development will continue on both LOMS and S2SM devices,
focusing on producing faster and more \mbox{resource-efficient} merge sorters for Versal Prime devices.
This development work may also lead to improved merge sorters even for Ultrascale+ products.
Since this merge sorting system is just being introduced here,
the system may enable significant improvements that have yet to be envisioned.

%% $$$$$$$$$$$$$$$$$$$$$$$$$$$$$$$$$$$$$$$$$$$$$$$$$$$$$$$$$$$$$$$$$$$$$$$$$$$$$

\appendices

\section{Design of a 3-way Merge Setup Array}

\label{app:3_way_merge_setup_design}

This appendix demonstrates a general methodology for designing a k-way List Offset k-column setup array.
The setup array is the initial 2-D array that is constructed in the hardware.

The general methodology is illustrated using the specific steps used to produce the 3c\_7r 3-way merge setup array,
previously shown in the left array in Section~\ref{sec:list_offset_3_way_merge}'s Fig.~\ref{fig:3c_7r_setup_and_outputs}.
The 3c\_7r 3-way LOMS merges 3 A, B, and C lists, each with 7 values.  Each list is sorted from the max at the
\_06 location to min at the \_00 location.

\begin{figure}[h]
  \centering
  \includegraphics[width=0.75\linewidth]{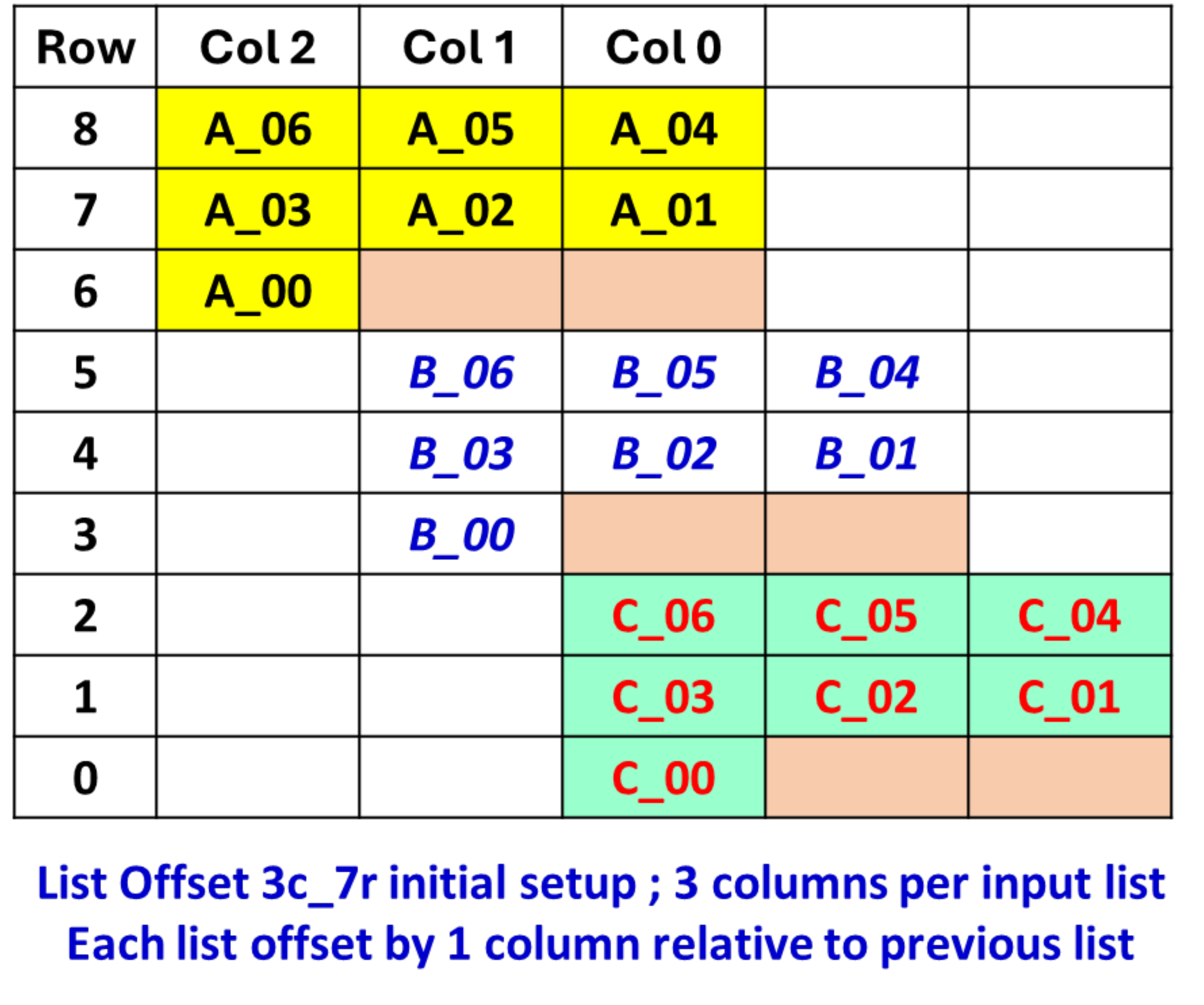}
  \caption{3-way Initial Setup: 3 lists, each with 7 values.}
  \label{fig:3c_7r_five_columns}
\end{figure}

\begin{figure}[h]
  \centering
  \includegraphics[width=1.0\linewidth]{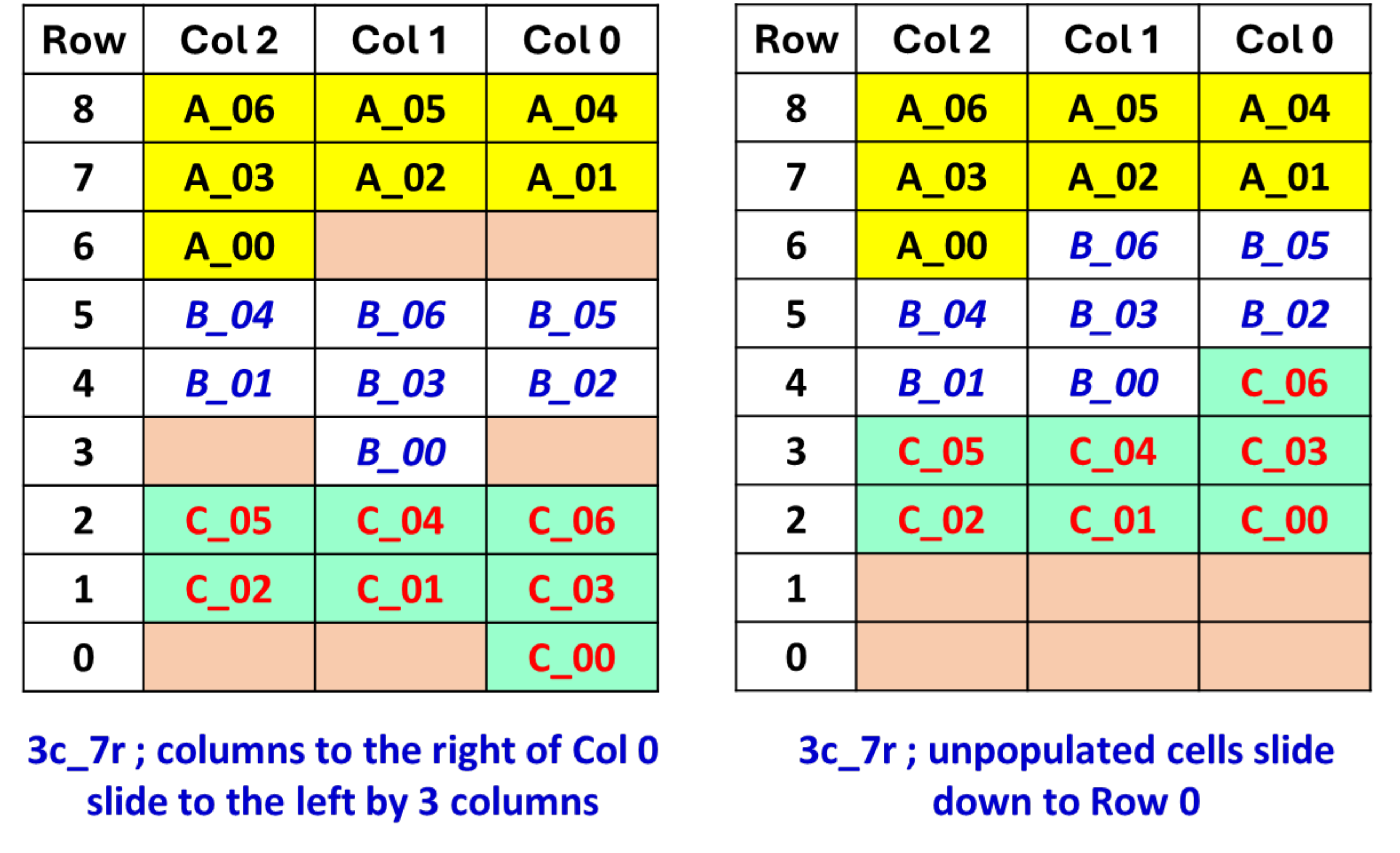}
  \caption{3-way Setup: After 3-column left slide, and vertical slide.}
  \label{fig:3c_7r_after_left_slide}
\end{figure}

Fig.~\ref{fig:3c_7r_five_columns} shows \mbox{how each} list is initially placed in a 2-D array.  Each list is offset by 1 column
from the previous list, and the resultant array has 5 columns.  As shown in the Fig.~\ref{fig:3c_7r_after_left_slide} left array,
the next setup step slides any data to the right of Col 0 3 columns to the left.
The data from all 3 input lists are now contained in 3 columns, as planned.

\begin{figure}[t]
  \centering
  \includegraphics[width=0.60\linewidth]{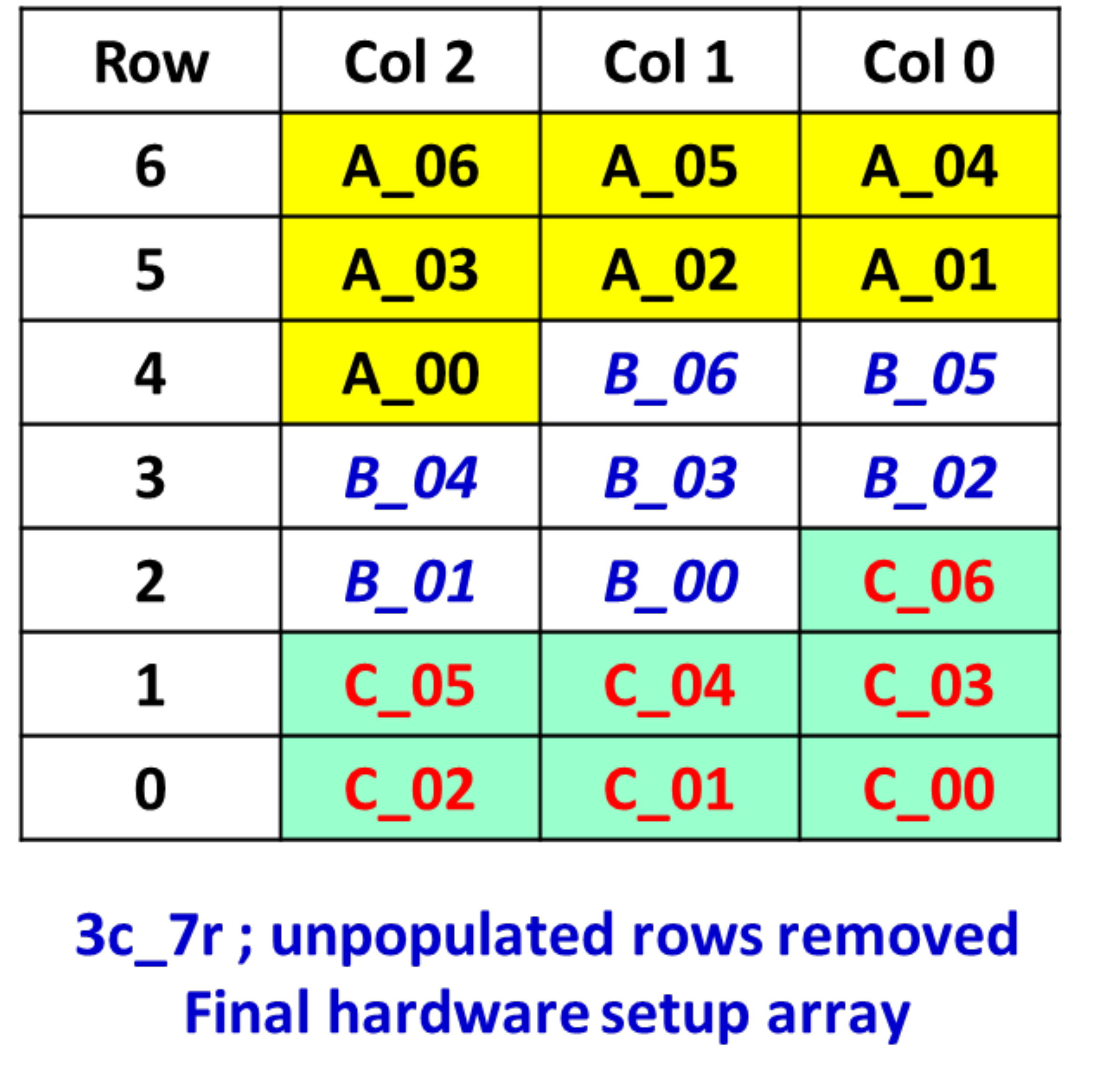}
  \caption{3-way Final Setup: After unpopulated rows removed.}
  \label{fig:3c_7r_unpopulated_removed_final_setup}
\end{figure}

The left Fig.~\ref{fig:3c_7r_after_left_slide} array contains several unpopulated cells.  The next step is to shift these unpopulated cells
to the bottom of the array, as was previously done for the \mbox{2-way} merge arrays in Section~\ref{sec:list_offset_2_way_merge}.
The right Fig.~\ref{fig:3c_7r_after_left_slide} array shows the array after the shift down of the unpopulated cells.

Any fully unpopulated rows are then removed, which was also done for \mbox{2-way} merge arrays in Section~\ref{sec:list_offset_2_way_merge}.
After removal of the fully unpopulated rows, the final setup array,
containing 7 fully populated rows, is shown in Fig.~\ref{fig:3c_7r_unpopulated_removed_final_setup}.
This is the setup array constructed in the hardware, and matches Fig.~\ref{fig:3c_7r_setup_and_outputs} in Section~\ref{sec:list_offset_3_way_merge}.
Section~\ref{sec:list_offset_3_way_merge} then uses a numeric example to demonstrate how the 3-way setup array is merge sorted.

As mentioned above, the methodology presented in this section can be used to produce a \mbox{k-column} setup array for any \mbox{k-way} LOMS.
This includes the \mbox{2-column} \mbox{2-way} merge sorters discussed in Section~\ref{sec:list_offset_2_way_merge}.
There were no figures similar to Fig.~\ref{fig:3c_7r_five_columns} in Section~\ref{sec:list_offset_2_way_merge},
followed by a \mbox{2-column} left slide, 
as it was assumed that the initial column offsets of the 2 sorted lists shown in the figures would be fairly obvious.

%% $$$$$$$$$$$$$$$$$$$$$$$$$$$$$$$$$$$$$$$$$$$$$$$$$$$$$$$$$$$$$$$$$$$$$$$$$$$$$

%% $$$$$$$$$$$$$$$$$$$$$$$$$$$$$$$$$$$$$$$$$$$$$$$$$$$$$$$$$$$$$$$$$$$$$$$$$$$$$

\EOD


\begin{thebibliography}{00}


\bibitem{Batcher_both} K. E. Batcher, ``Sorting Networks and Their Applications,''
\emph{Proceedings of the April 30--May 2, 1968, Spring Joint Computer Conference (AFIPS '68 Spring)},
pp. 307-314, 1968, ACM, https://doi.org/10.1145/1468075.1468121.

\bibitem{s2sm_Asilomar}
R. Kent and M. Pattichis,
``Stable Single-Stage 2-Way Merge Sorters'',
in \emph{2024 58th Asilomar Conference on Signals, Systems, and Computers},
Pacific Grove, CA, USA, 2024, pp. 270-277, doi: 10.1109/IEEECONF60004.2024.10942935. 

\bibitem{s2sm_patent}
R. Kent, 2025.
\emph{Single-stage merge sorting method and apparatus}. U.S. Patent 10,523,596.
US 12,306,776 B1

\bibitem{us_access_multiway_2022}
R. B. Kent and M. S. Pattichis,
``Design of High-Speed Multiway Merge Sorting Networks Using Fast Single-Stage N-Sorters and \mbox{N-filters},''
\emph{IEEE Access}., vol. 10, pp. 77980-77992, 2022, doi: 10.1109/ACCESS.2022.3193370.

\bibitem{multiway_merge_patent}
R. Kent and M. Pattichis, 2023.
\emph{Single-stage hardware sorting blocks and associated multiway merge sorting networks}.
U.S. Patent 12,306,776 B1.

%% 2012

\bibitem{Wisc_N_to_4} A. Farmahini-Farahani, HJ Duwe III, MJ Schulte, and K. Compton, ``Modular design of high-throughput, low-latency sorting units.'', \emph{IEEE Transactions on Computers}. 2012 May 29;62(7):1389-402.

\bibitem{Zuluaga_spiral} M. Zuluaga, P. Milder, and M. Püschel, 2012. ``Sorting Network IP Generator.'',
[Online], Available: http://www.spiral.net/hardware/sort/sort.html, Accessed on: Nov. 03, 2024

%% 2016

\bibitem{Zuluaga_streaming} M. Zuluaga, P. Milder, and M. Püschel, 2016. ``Streaming sorting networks.''
\emph{ACM Transactions on Design Automation of Electronic Systems (TODAES)}, 21(4), pp.1-30.

%% 2017

\bibitem{chen_prasanna} R. Chen and V. K. Prasanna,  (2017). ``Computer generation of high throughput and memory efficient sorting designs on FPGA.'', \emph{IEEE Transactions on Parallel and Distributed Systems}, 28(11), 3100-3113.

%% 2018

\bibitem{no_feedback} M. Saitoh, E.A. Elsayed, T. Van Chu, S. Mashimo, and K. Kise, 2018, April.
``A high-performance and cost-effective hardware merge sorter without feedback datapath.'' In \emph{2018 IEEE 26th Annual International Symposium on Field-Programmable Custom Computing Machines (FCCM)} (pp. 197-204). IEEE.

%% 2019

\bibitem{ferger_and_blott}
M. Ferger and M. Blott, Xilinx Inc, 2019. \emph{Circuits for and methods of merging streams of data to generate sorted output data}. U.S. Patent 10,523,596.

\bibitem{rths} A. Norollah, D. Derafshi, H. Beitollahi, and M. Fazeli. "RTHS: A low-cost high-performance real-time hardware sorter, using a multidimensional sorting algorithm." \emph{IEEE Transactions on Very Large Scale Integration (VLSI) Systems} 27, no. 7 (2019): 1601-1613.

%% 2022

\bibitem{flims} P. Papaphilippou, W. Luk, and C. Brooks, 2022.
``FLiMS: A fast lightweight 2-way merger for sorting.''
\emph{IEEE Transactions on Computers}, 71(12), pp.3215-3226.

%% 2023

\bibitem{Parallel_merge_sorter_patent_2023} 
Nie, Xiaoning, Mathias Kohlenz, and Jin-Soo Yoo.
``Parallel merge sorter circuit.''
U.S. Patent 11,803,509, issued October 31, 2023.

%% 2024

\bibitem{Oh_streaming_2024}
H. W. Oh, J. Park and S. E. Lee,
``DL-Sort: A Hybrid Approach to Scalable Hardware-Accelerated Fully-Streaming Sorting,''
in \emph{IEEE Transactions on Circuits and Systems II: Express Briefs}, March 2024, doi: 10.1109/TCSII.2024.3377255.

%% 2024

\bibitem{Ngai_2024} N. Ngai, I. Demertzis, J. Ghareh Chamani and D. Papadopoulos,
``Distributed \& Scalable Oblivious Sorting and Shuffling,''
in \emph{2024 IEEE Symposium on Security and Privacy (SP)}, San Francisco, CA, USA, 2024 pp. 152-152.
doi: 10.1109/SP54263.2024.00153

\bibitem{Adas_2007}
M.A. Adas and V. Sundarajan.
``VLSI architecture and implementation for single cycle insertion of multiple records into a priority sorted list.''
U.S. Patent 7,281,009, issued October 9, 2007.

\bibitem{sloping_and_shaking} Gao, Qingshi, and Zhiyong Liu. ``Sloping-and-shaking: Multiway merging and sorting.''
\emph{Science in China Series E: Technological Sciences 40}, (1997): 225-234.

\bibitem{median_filter_core_2024}
Sambamurthy, N. and Kamaraju, M., 2024. ``Scalable intelligent median filter core with adaptive impulse detector''. \emph{Analog Integrated Circuits and Signal Processing}, 118(3), pp.425-435.

\bibitem{us_1} R. B. Kent and M. S. Pattichis,
``Design, Implementation, and Analysis of High-Speed Single-Stage N-Sorters and N-Filters,''
\emph{IEEE Access}., vol. 9, pp. 2576-2591, Dec. 2020.

\bibitem{us_2} R. Kent and M. Pattichis,
``Use of Carry Chain Logic and Design System Extensions to Construct Significantly Faster and Larger Single-Stage \mbox{N-sorters} and \mbox{N-filters},''
\emph{IEEE Access}., vol. 10, pp. 79689-79702, Jul. 2022

\bibitem{us_3} R. Kent and M. Pattichis,
``Beyond 0-1: The 1-N Principle and Fast Validation of N-Sorter Sorting Networks,''
\emph{IEEE Access}., vol. 11, pp. 70574-70586, Jul 10. 2023




\end{thebibliography}
\end{document}